\documentstyle{article}
\begin{document}

\newcommand{\coment}[1]{ }

\thispagestyle{empty}

\begin{center}
\large
\phantom{xxx}
\vskip 2cm
\end{center}
\begin{flushright}
{\Large LAPTH--1249/08}
\end{flushright}
\vskip 1cm
\begin{center}
{\LARGE \bf L~a~n~H~E~P ---} \\[3mm]
{\Large a package for the automatic generation \\
of Feynman rules in field theory\\[5mm]
\bf Version 3.0} \\[1cm]
\vskip 1cm
{\Large A. V. Semenov}
\end{center}
\vskip 1cm
\begin{center}
Laboratory of Particle Physics,
     Joint Institute for Nuclear Research, \\
141 980 Dubna, Moscow Region, Russian Federation\\[4mm]
and \\
Laboratoire de Physique Th\'eorique LAPTH, Universit\'e de Savoie,\\
 Chemin de Bellevue, B.P. 110, F-74941 Annecy-le-Vieux, Cedex, France.\\[4mm]
\end{center}

\large


\vfill

{\bf Abstract}.  The LanHEP program version 3.0 for Feynman
rules generation from the Lagrangian is described. It reads
the Lagrangian written in a compact form, close to the one used in 
publications. It means that Lagrangian terms can be written 
with summation over indices of broken symmetries and using special symbols
for complicated expressions, such as covariant derivative and  
strength tensor for gauge fields. Supersymmetric theories can be described 
using the superpotential
    formalism and the 2-component fermion notation. The output
is Feynman rules in terms of physical fields and independent parameters
 in the form of CompHEP
model files, which allows one to start calculations of processes in
the new physical model. Alternatively, Feynman rules can be generated
in FeynArts format or as LaTeX table. One-loop counterterms can be 
generated in FeynArts format.

\vskip 1cm
\begin{center}
\begin{tabular}{rl}
{\it E-mail:} & {\tt semenov@jinr.ru}\\
{\it WWW page:} & { \tt http://theory.sinp.msu.ru/\~{}semenov/lanhep.html}
\end{tabular}
\end{center}

\vfill
\vfill
\begin{center}
{ \it 2008 }
\end{center}

\eject

\section*{Introduction}

LanHEP has been designed as part of the CompHEP package \cite{chep},
 a software for automatic calculations
in high energy physics. CompHEP allows symbolic computation
of the matrix element squared of any process with up to 6 incoming and
outgoing particles for a given physical model ({\it i.e.} a model defined by a
set of Feynman rules as 
a table of vertices in the momentum representation) and then numerical 
calculation of cross-sections and various distributions.

The main purpose of the new option given by LanHEP is to easily introduce 
new models.  Although this job is rather straightforward
and can be done manually, it requires careful calculations and in
modern theories with many particles and vertices, such as supersymmetric
models, can lead to errors and misprints. 

LanHEP makes possible the generation of
Feynman rules for propagators and vertices 
starting from the Lagrangian defined by a user in some simple format very
similar to canonical coordinate representation. It is possible to use
multiplets and substitution rules, for example for
 covariant derivative. Supersymmetric theories can be described using superpotential
    formalism and 2-component fermion notation.

The user should prepare a text file with description of all Lagrangian terms
 in the format close to the form used in standard publications. The output
 of the program is Feynman rules in CompHEP model tables, or in FeynArts
 \cite{FeynArts} format, or LaTeX table.

The LanHEP software is written in C programming language. The first
version \cite{lhepuser} was released in August 1996.

\section{Getting started with LanHEP}
\subsection{QED}

We start with a simple exercise, illustrating the
 main ideas and features of LanHEP. The first physical model is
Quantum Electrodynamics.

\begin{figure}[h]
\framebox{ \vbox{
\flushleft \tt \hspace*{1cm} model QED/1. \\
\hspace*{1cm} parameter ee=0.31333:'elementary electric charge'.\\
\hspace*{1cm} spinor e1/E1:(electron, mass me=0.000511). \\
\hspace*{1cm} vector A/A:(photon). \\
\hspace*{1cm} let F\^{}mu\^{}nu=deriv\^{}mu*A\^{}nu-deriv\^{}nu*A\^{}mu. \\
\hspace*{1cm} lterm -1/4*(F\^{}mu\^{}nu)**2 - 
		1/2*(deriv\^{}mu*A\^{}mu)**2. \\
\hspace*{1cm} lterm E1*(i*gamma*deriv+me)*e1. \\
\hspace*{1cm} lterm ee*E1*gamma*A*e1. }
}
\caption{LanHEP input file for the generation of QED Feynman rules}
\label{fig:qed1}
\end{figure}

The QED  Lagrangian is
$$ {\cal L}_{QED}=-\frac{1}{4}F_{\mu\nu} F^{\mu\nu} +
\bar e \gamma^\mu(i\partial_\mu +
g_e A_\mu)e - m\bar e e   $$
and the gauge fixing term in Feynman gauge has the form
$$ {\cal L}_{GF}=-\frac{1}{2}(\partial_\mu A^\mu)^2.$$
Here $e(x)$ is the spinor electron-positron field, $m$ is
 the electron mass, 
$A_\mu(x)$ is the vector photon field, 
$F^{\mu\nu}=\partial_\mu A^\nu-\partial_\nu A^\mu$,
  and $g_e$ is the elementary electric charge.

The LanHEP input file to generate the Feynman rules for QED is 
shown in Fig. \ref{fig:qed1}.


First of all, the input file consists of
statements. Each statement begins with one of the reserved keywords
 and ends
 by a full-stop '.' symbol.

First line says that this is a model with the name {\tt QED} and number 1.
This information is supplied for CompHEP, the name {\tt QED} will be 
displayed in its list of models.
In CompHEP package each model is described by four files:
 'varsN.mdl', 'funcN.mdl', 'prtclsN.mdl', 'lgrngnN.mdl' , where N is the
very number specified in the {\tt model} statement.

The {\tt model} statement stands first in the input file. If this statement
is absent, LanHEP does not generate the four standard CompHEP files, but just
 builds the model and prints
a diagnostic, if errors are found.

The second line in the input file contains declaration of the model parameter,
denoting elementary electric charge $g_e$ as {\tt ee}. For each parameter
 used in the model one should declare its
numeric value and an optional comment (it is also used in CompHEP menus).

The next two lines declare particles.
Statement names {\tt spinor, vector}
correspond to the particle spin. So, we declare electron 
 denoted by
{\tt e1} (the corresponding antiparticle name is {\tt E1}) and
photon denoted by {\tt A} (with antiparticle
name being  {\tt A}, since the antiparticle for photon is identical
to particle).

After the particle name we give in
brackets some
options. The first one is the full name of the particle, used in
CompHEP; the second option declares the
 mass of this particle.

The {\tt let} statement in the next  line declares the substitution rule
for symbol
{\tt F}.

Predefined name {\tt deriv}, which is reserved for the derivative 
$\frac{\partial} {\partial  x}$, will be replaced after 
the Fourier transformation by the momentum of the particle multiplied
 by $-i$.

The rest of the lines describe terms in the Lagrangian. Here the reserved name
{\tt gamma} denotes Dirac's $\gamma$-matrices.

One can see that the indices  are written separated with the
caret symbol '\^{}'.
Note that in the last two lines we have omitted indices. It means that
LanHEP  restores omitted indices automatically. Really, one can type
the last term in the full format: \begin{quote}
{\tt lterm ee*E1\^{}a*gamma\^{}a\^{}b\^{}mu*A\^{}mu*e1\^{}b}. \end{quote}
It corresponds to $g_e \bar e_a \gamma^\mu_{ab} e_b A_\mu$ 
with all indices written. Note that the order of objects
 in the monomial is important to restore indices automatically.

\subsection{QCD}

Now let us consider the case of Quantum
Chromodynamics. The Lagrangian for the gluon fields reads

$$ L_{YM} = -\frac{1}{4}F^{a\mu\nu}F^a_{\mu\nu},$$
where
$$
F^a_{\mu\nu}=\partial_\mu G^a_\nu-\partial_\nu G^a_\mu-
	g_s f^{abc}G^b_\mu G^c_\nu,$$
$G^a_\mu(x)$ is the gluon field, $g_s$ is the strong coupling constant
 and $f^{abc}$ are purely imaginary structure
constants of the $SU(3)$ color group.

The quark kinetic term and its interaction  with the gluon has the form
$$ L_F = \bar q_i \gamma^\mu \partial_\mu q_i + g_s  \lambda^a_{ij} 
\bar q_i\gamma^\mu  q_j G_\mu^c,$$
where $\lambda^a_{ij}$ are Gell-Mann matrices.

Gauge fixing terms in Feynman gauge together with the corresponding 
Faddev-Popov ghost term  are
$$ -\frac{1}{2}(\partial_\mu G^\mu_a)^2 + ig_s f^{abc} \bar c^a G^b_\mu
\partial^\mu c^c,$$
where $(c, \bar c)$ are unphysical ghost fields.

The corresponding LanHEP input file is shown in Fig. \ref{fig:qcd1}.

\begin{figure}[th]
\framebox{ \vbox{
\flushleft \tt \hspace*{1cm} model QCD/2. \\
\hspace*{1cm} parameter gg=1.117:'Strong coupling'.\\
\hspace*{1cm} spinor q/Q:(quark, mass mq=0.01, color c3). \\
\hspace*{1cm} vector G/G:(gluon, color c8, gauge). \\
\hspace*{1cm} let F\^{}mu\^{}nu\^{}a = deriv\^{}nu*G\^{}mu\^{}a -
           deriv\^{}mu*G\^{}nu\^{}a -
\hspace*{2cm} gg*f\_SU3\^{}a\^{}b\^{}c*G\^{}mu\^{}b*G\^{}nu\^{}c.\\
\hspace*{1cm} lterm -F**2/4-(deriv*G)**2/2.\\
\hspace*{1cm} lterm Q*(i*gamma*deriv+mq)*q.\\
\hspace*{1cm} lterm i*gg*f\_SU3*ccghost(G)*G*deriv*ghost(G). \\
\hspace*{1cm} lterm gg*Q*gamma*lambda*G*q. }
}
\caption{Input file for the generation of QCD Feynman rules}
\label{fig:qcd1}
\end{figure}

\begin{table}[ht]
\caption{QCD Feynman rules generated by LanHEP in LaTeX output format}
\begin{center}
\begin{tabular}{|llll|l|} \hline
\multicolumn{4}{|c|}{Fields in the vertex} & Variational derivative
 of Lagrangian by fields \\ \hline
${G}_{\mu p }$ & ${\bar\eta^G}_{q }$ & ${\eta^G}_{r }$ &  & $- g_s p_3^\mu
 f_{p q r} $\\[2mm]
${\bar q}_{a p }$ & ${q}_{b q }$ & ${G}_{\mu r }$ &  & $ g_s
 \gamma_{a b}^\mu \lambda_{p q}^r $\\[2mm]
${G}_{\mu p }$ & ${G}_{\nu q }$ & ${G}_{\rho r }$ &  & $ g_s f_{p q r}
 \big(p_3^\nu g^{\mu \rho} -p_2^\rho g^{\mu \nu} -p_3^\mu g^{\nu \rho}
 +p_1^\rho g^{\mu \nu} +p_2^\mu g^{\nu \rho} -p_1^\nu g^{\mu \rho}
 \big)$\\[2mm]
${G}_{\mu p }$ & ${G}_{\nu q }$ & ${G}_{\rho r }$ & ${G}_{\sigma s }$
 & $ g_s^2 \big(g^{\mu \rho} g^{\nu \sigma} f_{p q t} f_{r s t} -g^{\mu
 \sigma} g^{\nu \rho} f_{p q t} f_{r s t} +g^{\mu \nu} g^{\rho \sigma}
 f_{p r t} f_{q s t} $ \\[2mm]
 & & & & $+g^{\mu \nu} g^{\rho \sigma} f_{p s t} f_{q r t} -g^{\mu
 \sigma} g^{\nu \rho} f_{p r t} f_{q s t} -g^{\mu \rho} g^{\nu \sigma}
 f_{p s t} f_{q r t} \big)$\\ \hline
\end{tabular}
\end{center}
\end{table}

Since QCD uses objects with
color indices, one has to declare the indices of these objects.
There are three types of color indices supported by LanHEP. These types 
are referred as {\tt color c3} (color triplets), {\tt color c3b}
 (color antitriplets), and {\tt color c8} (color octets). 
One can see that {\tt color c3} index type appears among the options in
the quark {\tt q} declaration, and the {\tt color c8} one in the gluon
{\tt G} declaration. Antiquark {\tt Q} has got color index of type {\tt 
color c3b} as antiparticle to quark.
 LanHEP allows  contraction of an index of type {\tt color c3} only with
another index of type {\tt color c3b}, and two indices of type {\tt color
c8}. Of course, in Lagrangian terms each index
has to be contracted with its partner, since the Lagrangian has to be a scalar.

LanHEP allows also to use in the Lagrangian terms a predefined symbol 
{\tt lambda} with the three indices of types {\tt color c3, color c3b,
color c8} corresponding to Gell-Mann  
$\lambda$-matrices. Symbol {\tt f\_SU3} denotes the structure constant
$f^{abc}$ of color $SU(3)$
group (all three indices have the type {\tt color c8}).

Option {\tt gauge}
in the declaration of {\tt G} allows to use names {\tt ghost(G)} and {\tt
ccghost(G)} for the ghost fields $c$ and $\bar c$ in Lagrangian terms
and in {\tt let} statements. 

Table 1 shows Feynman rules generated by LanHEP in LaTeX format 
after processing the input file presented in Fig. \ref{fig:qcd1}. 
Four gluon  vertex rule is indicated in the last line. Note that the
output in CompHEP format has no 4-gluon vertex explicitly; it is 
expressed effectively through 3-leg vertices by a constant propagator
 of some auxiliary field (see section \ref{imp} for more details).

\subsection{Higgs sector of the Standard Model}

\begin{figure}[th]
\framebox{ \vbox{ \hsize 15cm
\flushleft \tt 
\hspace*{1cm}model Higgs/1.\\
\hspace*{1cm}parameter  EE  = 0.31333 : 'Electromagnetic coupling constant',\\
\hspace*{2cm}	   SW  = 0.4740  : 'sin of the Weinberg angle (PDG-94)',\\
\hspace*{2cm}           CW  = Sqrt(1-SW**2) : 'cos of the Weinberg angle'.\\
\hspace*{1cm}let        g=EE/SW, g1=EE/CW.\\[2mm]
\hspace*{1cm}vector  A/A: (photon, gauge),\\
\hspace*{2cm}	Z/Z:('Z boson', mass MZ = 91.187,  gauge),\\
\hspace*{2cm}	'W+'/'W-': ('W boson', mass MW = MZ*CW, gauge).\\
\hspace*{1cm}scalar H/H:(Higgs, mass MH = 200, width wH = 1.461).\\[2mm]
\hspace*{1cm}let B = -SW*Z+CW*A.\\
\hspace*{1cm}let W = \{'W+', CW*Z+SW*A, 'W-'\}.\\
\hspace*{1cm}let phi = \{ -i*gsb('W+'),  (vev(2*MW/EE*SW)+H+i*gsb(Z))/Sqrt2 \},\\
\hspace*{2cm}    Phi = anti(phi).\\[2mm]
\hspace*{1cm}lterm -2*lambda*(phi*anti(phi)-v**2/2)**2  where\\
\hspace*{3cm}	lambda=(g*MH/MW)**2/16, v=2*MW*SW/EE.\\
\hspace*{1cm}let D\^{}a\^{}b\^{}mu =(deriv\^{}mu+i*g1/2*B\^{}mu)*delta(2)\^{}a\^{}b\\
\hspace*{3cm}        +i*g/2*taupm\^{}a\^{}b\^{}c*W\^{}mu\^{}c,\\
\hspace*{2cm}    Dc\^{}a\^{}b\^{}mu=(deriv\^{}mu-i*g1/2*B\^{}mu)*delta(2)\^{}a\^{}b\\
\hspace*{3cm}        -i*g/2*taupm\^{}a\^{}b\^{}c*anti(W)\^{}mu\^{}c.\\
\hspace*{1cm}lterm D\^{}a\^{}b\^{}mu*phi\^{}b*Dc\^{}a\^{}c\^{}mu*Phi\^{}c. }
}
\caption{Input file for the Higgs sector of the Standard Model}
\label{fig:hs1}
\end{figure}

The third example illustrates using the multiplets
in the framework of LanHEP. Let us consider the 
Higgs sector of the Standard Model. The Higgs doublet
can be defined as $$\Phi = \left(
\begin{array}{c}
-iW^+_f\\
(\frac{2M_W}{es_s}+H+iZ_f)/\sqrt{2}\end{array}\right),$$
where $s_w$ is the sinus of the weak angle, $H$ is the Higgs
field, $Z_f$ and $W^\pm_f$ are goldstone bosons corresponding
to $Z$ and $W^\pm$ gauge fields. The  Higgs potential
 reads as $${\cal L}_H= -2\lambda(\Phi\Phi^*-\upsilon^2/2)^2,$$
where $\lambda=(gM_H/M_W)^2/16$, and the vaccuum expectation value
$\upsilon=2M_W/g$ (here and below $g=e/s_w$, $g'=e/c_w$.)

The gauge interaction of a Higgs doublet is given by the term
$(D_\mu\Phi)(D^\mu\Phi)^*$, where 
$$ D_\mu = \partial_\mu +ig'B_\mu/2 + ig\vec \tau \vec W_\mu.$$
Here $B$ is $U(1)$ singlet and $W$ is $SU(2)$ triplet,
$$B_\mu=-s_wZ_\mu+c_wA_\mu,\;\;\;\;\;
  W_\mu^a=\left( \begin{array}{c}
  W^+_\mu\\
  c_wZ_\mu+s_wA_\mu\\
  W^-_\mu \end{array}\right),$$
and $\vec\tau$ is the vector $(\tau^+, \tau^3, \tau^-)$.

The above model can be represented by the LanHEP code shown in Table \ref{fig:hs1}.

New features in this example include the definition of special
symbols for coupling constants $g$, $g'$, gauge and Higgs
fields, and the covariant derivative by means of {\tt let} statement. 
Note that in the declaration
for the fields we have used dummy indices (vector and isospin).
The multiplets are defined by the components in the curly brackets.

Note that the option {\tt gauge}
in the declaration of gauge fields allows to use the name {\tt gsb(Z)} and {\tt
gsb('W+')} for the goldstone bosons. 

\begin{table}
\caption{LanHEP output: Feynman rules for Higgs self-interaction}
\label{fig:hs2}
\begin{center}
\begin{tabular}{|l|l|} \hline
Fields in the vertex & Variational derivative of Lagrangian by fields \\ \hline
${H}_{}$ \phantom{-} ${H}_{}$ \phantom{-} ${H}_{}$ \phantom{-}  &
	$-\frac{3}{2}\frac{ e MH{}^2 }{ M_W s_w}$\\[2mm]
${H}_{}$ \phantom{-} $W^+_F{}_{}$ \phantom{-} $W^-_F{}_{}$ \phantom{-}  &
	$-\frac{1}{2}\frac{ e MH{}^2 }{ M_W s_w}$\\[2mm]
${H}_{}$ \phantom{-} $Z_F{}_{}$ \phantom{-} $Z_F{}_{}$ \phantom{-}  &
	$-\frac{1}{2}\frac{ e MH{}^2 }{ M_W s_w}$\\[2mm]
${H}_{}$ \phantom{-} ${H}_{}$ \phantom{-} ${H}_{}$ \phantom{-} ${H}_{}$ \phantom{-}  &
	$-\frac{3}{4}\frac{ e{}^2  MH{}^2 }{ M_W{}^2  s_w{}^2 }$\\[2mm]
${H}_{}$ \phantom{-} ${H}_{}$ \phantom{-} $W^+_F{}_{}$ \phantom{-} $W^-_F{}_{}$ \phantom{-}  &
	$-\frac{1}{4}\frac{ e{}^2  MH{}^2 }{ M_W{}^2  s_w{}^2 }$\\[2mm]
${H}_{}$ \phantom{-} ${H}_{}$ \phantom{-} $Z_F{}_{}$ \phantom{-} $Z_F{}_{}$ \phantom{-}  &
	$-\frac{1}{4}\frac{ e{}^2  MH{}^2 }{ M_W{}^2  s_w{}^2 }$\\[2mm]
$W^+_F{}_{}$ \phantom{-} $W^+_F{}_{}$ \phantom{-} $W^-_F{}_{}$ \phantom{-} $W^-_F{}_{}$ \phantom{-}  &
	$-\frac{1}{2}\frac{ e{}^2  MH{}^2 }{ M_W{}^2  s_w{}^2 }$\\[2mm]
$W^+_F{}_{}$ \phantom{-} $W^-_F{}_{}$ \phantom{-} $Z_F{}_{}$ \phantom{-} $Z_F{}_{}$ \phantom{-}  &
	$-\frac{1}{4}\frac{ e{}^2  MH{}^2 }{ M_W{}^2  s_w{}^2 }$\\[2mm]
$Z_F{}_{}$ \phantom{-} $Z_F{}_{}$ \phantom{-} $Z_F{}_{}$ \phantom{-} $Z_F{}_{}$ \phantom{-}  &
	$-\frac{3}{4}\frac{ e{}^2  MH{}^2 }{ M_W{}^2  s_w{}^2 }$\\ \hline
\end{tabular}
\end{center}
\end{table}

\begin{table}
\caption{LanHEP output: Feynman rules for Higgs gauge interaction}
\label{fig:hs3}
\begin{center}
\begin{tabular}{|l|l|} \hline
Fields in the vertex & Variational derivative of Lagrangian by fields \\ \hline
${A}_{\mu }$ \phantom{-} $W^+{}_{\nu }$ \phantom{-} $W^-_F{}_{}$ \phantom{-}  &
	$ i e M_Wg^{\mu \nu} $\\[2mm]
${A}_{\mu }$ \phantom{-} $W^+_F{}_{}$ \phantom{-} $W^-{}_{\nu }$ \phantom{-}  &
	$- i e M_Wg^{\mu \nu} $\\[2mm]
${A}_{\mu }$ \phantom{-} $W^+_F{}_{}$ \phantom{-} $W^-_F{}_{}$ \phantom{-}  &
	$- e\big(p_2^\mu -p_3^\mu \big)$\\[2mm]
${H}_{}$ \phantom{-} $W^+{}_{\mu }$ \phantom{-} $W^-{}_{\nu }$ \phantom{-}  &
	$\frac{ e M_W}{ s_w}g^{\mu \nu} $\\[2mm]
${H}_{}$ \phantom{-} $W^+{}_{\mu }$ \phantom{-} $W^-_F{}_{}$ \phantom{-}  &
	$-\frac{1}{2}\frac{ i e}{ s_w}\big(p_1^\mu -p_3^\mu \big)$\\[2mm]
${H}_{}$ \phantom{-} $W^+_F{}_{}$ \phantom{-} $W^-{}_{\mu }$ \phantom{-}  &
	$\frac{1}{2}\frac{ i e}{ s_w}\big(p_2^\mu -p_1^\mu \big)$\\[2mm]
${H}_{}$ \phantom{-} ${Z}_{\mu }$ \phantom{-} ${Z}_{\nu }$ \phantom{-}  &
	$\frac{ e M_W}{ c_w{}^2  s_w}g^{\mu \nu} $\\[2mm]
${H}_{}$ \phantom{-} ${Z}_{\mu }$ \phantom{-} $Z_F{}_{}$ \phantom{-}  &
	$-\frac{1}{2}\frac{ i e}{ c_w s_w}\big(p_1^\mu -p_3^\mu \big)$\\[2mm]
$W^+{}_{\mu }$ \phantom{-} $W^-_F{}_{}$ \phantom{-} ${Z}_{\nu }$ \phantom{-}  &
	$-\frac{ i e M_W s_w}{ c_w}g^{\mu \nu} $\\[2mm]
$W^+{}_{\mu }$ \phantom{-} $W^-_F{}_{}$ \phantom{-} $Z_F{}_{}$ \phantom{-}  &
	$\frac{1}{2}\frac{ e}{ s_w}\big(p_3^\mu -p_2^\mu \big)$\\[2mm]
$W^+_F{}_{}$ \phantom{-} $W^-{}_{\mu }$ \phantom{-} ${Z}_{\nu }$ \phantom{-}  &
	$\frac{ i e M_W s_w}{ c_w}g^{\mu \nu} $\\[2mm]
$W^+_F{}_{}$ \phantom{-} $W^-{}_{\mu }$ \phantom{-} $Z_F{}_{}$ \phantom{-}  &
	$\frac{1}{2}\frac{ e}{ s_w}\big(p_1^\mu -p_3^\mu \big)$\\[2mm]
$W^+_F{}_{}$ \phantom{-} $W^-_F{}_{}$ \phantom{-} ${Z}_{\mu }$ \phantom{-}  &
	$-\frac{1}{2}\frac{ (1-2 s_w {}^2) e}{ c_w s_w}\big(p_1^\mu -p_2^\mu \big)$\\[2mm]
${A}_{\mu }$ \phantom{-} ${A}_{\nu }$ \phantom{-} $W^+_F{}_{}$ \phantom{-} $W^-_F{}_{}$ \phantom{-}  &
	$2 e{}^2 g^{\mu \nu} $\\[2mm]
${A}_{\mu }$ \phantom{-} ${H}_{}$ \phantom{-} $W^+{}_{\nu }$ \phantom{-} $W^-_F{}_{}$ \phantom{-}  &
	$\frac{1}{2}\frac{ i e{}^2 }{ s_w}g^{\mu \nu} $\\[2mm]
${A}_{\mu }$ \phantom{-} ${H}_{}$ \phantom{-} $W^+_F{}_{}$ \phantom{-} $W^-{}_{\nu }$ \phantom{-}  &
	$-\frac{1}{2}\frac{ i e{}^2 }{ s_w}g^{\mu \nu} $\\[2mm]
${A}_{\mu }$ \phantom{-} $W^+{}_{\nu }$ \phantom{-} $W^-_F{}_{}$ \phantom{-} $Z_F{}_{}$ \phantom{-}  &
	$-\frac{1}{2}\frac{ e{}^2 }{ s_w}g^{\mu \nu} $\\[2mm]
${A}_{\mu }$ \phantom{-} $W^+_F{}_{}$ \phantom{-} $W^-{}_{\nu }$ \phantom{-} $Z_F{}_{}$ \phantom{-}  &
	$-\frac{1}{2}\frac{ e{}^2 }{ s_w}g^{\mu \nu} $\\[2mm]
${A}_{\mu }$ \phantom{-} $W^+_F{}_{}$ \phantom{-} $W^-_F{}_{}$ \phantom{-} ${Z}_{\nu }$ \phantom{-}  &
	$\frac{ (1-2 s_w {}^2) e{}^2 }{ c_w s_w}g^{\mu \nu} $\\[2mm]
${H}_{}$ \phantom{-} ${H}_{}$ \phantom{-} $W^+{}_{\mu }$ \phantom{-} $W^-{}_{\nu }$ \phantom{-}  &
	$\frac{1}{2}\frac{ e{}^2 }{ s_w{}^2 }g^{\mu \nu} $\\[2mm]
${H}_{}$ \phantom{-} ${H}_{}$ \phantom{-} ${Z}_{\mu }$ \phantom{-} ${Z}_{\nu }$ \phantom{-}  &
	$\frac{1}{2}\frac{ e{}^2 }{ c_w{}^2  s_w{}^2 }g^{\mu \nu} $\\[2mm]
${H}_{}$ \phantom{-} $W^+{}_{\mu }$ \phantom{-} $W^-_F{}_{}$ \phantom{-} ${Z}_{\nu }$ \phantom{-}  &
	$-\frac{1}{2}\frac{ i e{}^2 }{ c_w}g^{\mu \nu} $\\[2mm]
${H}_{}$ \phantom{-} $W^+_F{}_{}$ \phantom{-} $W^-{}_{\mu }$ \phantom{-} ${Z}_{\nu }$ \phantom{-}  &
	$\frac{1}{2}\frac{ i e{}^2 }{ c_w}g^{\mu \nu} $\\[2mm]
$W^+{}_{\mu }$ \phantom{-} $W^+_F{}_{}$ \phantom{-} $W^-{}_{\nu }$ \phantom{-} $W^-_F{}_{}$ \phantom{-}  &
	$\frac{1}{2}\frac{ e{}^2 }{ s_w{}^2 }g^{\mu \nu} $\\[2mm]
$W^+{}_{\mu }$ \phantom{-} $W^-{}_{\nu }$ \phantom{-} $Z_F{}_{}$ \phantom{-} $Z_F{}_{}$ \phantom{-}  &
	$\frac{1}{2}\frac{ e{}^2 }{ s_w{}^2 }g^{\mu \nu} $\\[2mm]
$W^+{}_{\mu }$ \phantom{-} $W^-_F{}_{}$ \phantom{-} ${Z}_{\nu }$ \phantom{-} $Z_F{}_{}$ \phantom{-}  &
	$\frac{1}{2}\frac{ e{}^2 }{ c_w}g^{\mu \nu} $\\[2mm]
$W^+_F{}_{}$ \phantom{-} $W^-{}_{\mu }$ \phantom{-} ${Z}_{\nu }$ \phantom{-} $Z_F{}_{}$ \phantom{-}  &
	$\frac{1}{2}\frac{ e{}^2 }{ c_w}g^{\mu \nu} $\\[2mm]
$W^+_F{}_{}$ \phantom{-} $W^-_F{}_{}$ \phantom{-} ${Z}_{\mu }$ \phantom{-} ${Z}_{\nu }$ \phantom{-}  &
	$\frac{1}{2}\frac{ (1-2 s_w {}^2){}^2  e{}^2 }{ c_w{}^2  s_w{}^2 }g^{\mu \nu} $\\[2mm]
${Z}_{\mu }$ \phantom{-} ${Z}_{\nu }$ \phantom{-} $Z_F{}_{}$ \phantom{-} $Z_F{}_{}$ \phantom{-}  &
	$\frac{1}{2}\frac{ e{}^2 }{ c_w{}^2  s_w{}^2 }g^{\mu \nu} $\\ \hline
\end{tabular}
\end{center}
\end{table}


\section{Structure of LanHEP input file}

The LanHEP input file is the sequence of statements,
each starts with a special identifier (such as {\tt parameter,
lterm} etc) and ends with the full-stop '{\bf .}' symbol. 
Statement can occupy several lines in the input file.

This section is aimed to clarify the syntax of LanHEP input files, {\it i.e.} 
the structure of the statements.

\subsection{Constants and identifiers}

First of all, each word in any statement is either an {\it identifier}
or a {\it constant}.

 Indentfiers are the names of particles, parameters etc.
Examples of identifiers from the previous section are particle names 
\begin{quote} {\tt e1 E1 A q Q G} \end{quote}
 The first word in each statement is also
an identifier, defining the function which this statement
performs. The identifiers are usually combinations of letters and digits
starting with a letter. If an identifier does not respect this 
rule, it should be quoted. For example, the names of $W^\pm$ bosons
must be written as {\tt 'W+'} and {\tt 'W-'}, since they contain 
'+' and '--' symbols.

Constants can be classified  as
\begin{itemize}
\item {\it integers}: they consist of optional sign followed by one or
more decimal digits, such as  
\begin{quote} { \tt 0 \, 1 \, -1 \, 123 \, -98765} \end{quote}
Integers can  appear in Lagrangian terms, parameter definition and
in other expressions.
\item {\it Floating point numbers} 
 include optional sign, several decimal digits of
mantissa with an embedded period (decimal point) with at least one digit
before and after the period, and optional exponent. The exponent, if
present, consist of letter {\tt E} or {\tt e} followed by an optional sign
and one or more decimal digits. Valid examples of floating point
 numbers are 
\begin{quote} {\tt 1.0 \, -1.0 \, 0.000511 \, 5.11e-4} \end{quote}
Floating point numbers are used only as parameter values (coupling
constants, particle masses etc). They can not
be explicitly used in Lagrangian terms.
\item {\it String constants} may include arbitrary symbols. They are
used as comments in parameter statements, full particle
names in the declaration of a particle, etc. Examples from the previous
section are 
\begin{quote} {\tt electron \\ photon} \end{quote} 
If a string constant
contains any character besides letters and digits or does not begin 
with a letter, it should be quoted. For example, the comments in QED and
 QCD input files (see previous section)
 contain blank spaces, so they are quoted:
\begin{quote} {\tt 'elementary electric charge' \\ 'Strong coupling'}
\end{quote}
\end{itemize}

\subsection{Comments}

User can include comments into the LanHEP input file in two ways. First,
symbol {\tt '\%'} denotes the comment till the end of current line.
Second way allows one to comment any number of lines by putting a part of
input file  between
{\tt '/*'} (begin of comment) and {\tt '*/'} (end of comment) symbols.

\subsection{Including files}

LanHEP allows the user to divide the input file into several files.
 To include the file {\sl file}, the user should
use the statement \begin{quote} {\tt read \sl file\tt . } \end{quote}
 The standard extension '{\tt .mdl}' of
the file name may be omitted in this statement.

Another way to include a file is provided by the {\tt use} statement as
\begin{quote} {\tt use  \sl file\tt . } \end{quote}
 The {\tt use} statement
reads the {\tt file} only once, next appearances of this statement with the
same argument do nothing. This function prevents multiple reading of the
 same file.
This form can be used mainly to include
some standard modules, such as declaration of Standard Model particles
to be used for writing some extensions of this model.

\subsection{Conditional processing of the model}

Let us consider the LanHEP input files for the Standard Model with both the 
t'Hooft-Feynman and unitary gauges. It is clear that these input files differ
 by only a few
lines --- the declaration of gauge bosons and Higgs doublet. It is more
convenient to have only one model definition file. In this case the 
conditional statements are necessary. LanHEP allows the user to define
several {\it keys\/}, and use these keys to branch among several variants
of the model. The keys have to be declared by the {\tt keys} statement:
\begin{quote}{\tt keys \sl name1=value1, name2=value2, ... \tt .}\end{quote}
Then one can use the conditional statements:
\begin{quote}{\tt
do\_if \sl key\tt ==\sl value1.\\
\hspace*{1cm} actions1\\
\tt do\_else\_if \sl key\tt ==\sl value2.\\
\hspace*{1cm} actions2\\
\tt do\_else\_if \sl key\tt ==\sl value3.\\
\hspace*{1cm} ... \\
\tt do\_else. \\
\hspace*{1cm} \sl default actions\\
\tt end\_if. } \end{quote}

The statements {\tt do\_else\_if} and {\tt do\_else} are not mandatory.
The value of the key is a number or symbolic string.
An example of using these statements in the Standard model may read:
\begin{quote}{\tt
keys Gauge=unitary.\\[4mm]
do\_if Gauge==Feynman.\\
vector Z/Z:('Z-boson',mass MZ = 91.187, width wZ = 2.502, gauge).\\
do\_else\_if Gauge==unitary.\\
vector Z/Z:('Z-boson',mass MZ = 91.187, width wZ = 2.502).\\
do\_else.\\
write('Error: key Gauge must be either Feynman or unitary').\\
quit.\\
end\_if.} \end{quote}

Thus, to change the choice of gauge fixing in the generated Feynman rules
it is enough to modify one word in the input file. There is another 
way, to set the key value from the command line at the launch of LanHEP:
\begin{quote}{\tt
 lhep -key Gauge=Feynman \sl filename.}\end{quote}
 
If the value of key is set from the command line at the program launch,
the value for this key in the {\tt keys} statement is ignored.

There is also a short form of {\tt do\_if} statement:
\begin{quote}{\tt
do\_if(\sl key\tt ==\sl value, (statement1), (statement2)).} \end{quote}
If the {\sl key} is defined equal to {\sl value}, then the {\sl statement1}
is processed. Otherwise the program executes {\sl statement2}. The form
with only one {\sl statement1} is allowed. An example could be:
\begin{quote}{\tt
do\_if(electron\_mass==on, (parameter Me=0.511e-3), (let Me=0)).}\end{quote}


\section{Objects in the expressions for Lagrangian terms}

Each symbol which may appear in algebraic expressions (names of
parameters, fields, etc) has a fixed order of indices and their types. 
If this object is used in any expression, one should write its indices in
 the same order as they were defined when the object was declared.

Besides the types of indices corresponding to the color $SU(3)$ group: {\tt color
 c3}
 (color triplet), {\tt color
c3b} (color antitriplet) and {\tt color c8} (color octet)
described in the previous example, there are default types 
of indices for Lorentz group:  {\tt vector}, {\tt spinor} and {\tt cspinor}
(antispinor). User can also declare
new types of indices corresponding to the 
symmetries other than color $SU(3)$ group. In this case any object
(say, particle)
may have indices related to this new group. This possibility will be
described in Section \ref{newgrp}.

If an index appears twice in some monomial of an expression, LanHEP
assumes summation over this index. Types of such indices must allow
the contraction, {\it i.e.} they should be one of the pairs:
{\tt spinor} and {\tt cspinor}, two {\tt vector}, {\tt color c3} and
{\tt color c3b}, two {\tt color c8}.  

In general the following objects are available to appear in the expressions
for a Lagrangian: integers and identifiers of parameters, particles,
 specials,
let-substitutions and arrays.

There are also predefined symbols {\tt i}, denoting imaginary unit $i$
($i^2=-1$) and {\tt Sqrt2}, which is a parameter with value
$\sqrt{2}$.

\subsection{Parameters}

{\it Parameters\/} are scalar objects ({\it i.e.} they have no indices).
Parameters denote coupling constants,
masses and widths of particles, etc. To introduce a new parameter 
 one should use the  {\tt parameter} statement,
which  has the generic form
\begin{quote} { \tt parameter \sl name\/\tt =\sl value\/\tt :\sl comment. }
\end{quote}
\begin{itemize}
\item \underline{\sl name} is an identifier of newly created parameter. 
\item \underline{\sl value} is an integer or floating point number or an
expression. One can use previously declared parameters
and integers joined by standard arithmetical operators {\tt '+', '-',
 '*', '/'}, and {\tt '**'} (power).  
\item \underline{\sl comment }
is an optional comment to clarify the meaning of parameter,
it is used in CompHEP help windows. 
Comment has to be a string constant, so if it contains blank spaces
or other special characters, it must be quoted (see Section 3). 
\end{itemize}

In the expression for parameters one can use functions {\tt sqrt, pow, 
sin, asin, cos, acos, tan, atan, atan2, fabs\/}  
(same as in C programming language). One can add new functions through the
use of the statement
\begin{quote} { \tt external\_func(\sl function, arity\tt)}.
\end{quote}

{\sl Arity\/} is the number of arguments of the function. If this
function is absent in the standard C library, it should be written
in C and added to the library file {\tt usrlib.a\/} in the CompHEP
working directory.

\subsection{Particles}

{\it Particles\/ } are objects to denote physical particles. 
They may have indices. It is possible to use five statements to declare
a new particle, at the same time statements define
the corresponding Lorentz index(es):
 \begin{quote}
{ \tt  scalar \sl P/aP\tt :(\sl options\tt ).} \\
{ \tt  spinor \sl P/aP\tt :(\sl options\tt ). } \\
{ \tt  vector \sl P/aP\tt :(\sl options\tt ). } \\
{ \tt  spinor3 \sl P/aP\tt :(\sl options\tt ). } \\
{ \tt  tensor \sl P/aP\tt :(\sl options\tt ). }   \end{quote}
{\tt spinor3} stands for spin $3/2$, {\sl P} and {\sl aP} are identifiers of particle and antiparticle.
In the case of truly neutral particles (when antiparticle is identical to 
the particle itself) one should use the form {\sl P/P} with identical
 names for particle and antiparticle. 

It is possible to write only the particle name, e.g. 
\begin{quote} {\tt  scalar  \sl P\tt :(\sl options\tt )}. \end{quote}
 In this case the name of corresponding
antiparticle is generated automatically. It satisfies the usual CompHEP
convention, when the name of antiparticle differs from particle by 
altering the case of the first letter. 
So for electron name {\tt e1} automatically generated antiparticle
name will be {\tt E1}. If the name contains symbol '+' it is replaced by
'--' and vice versa.

The {\sl option} is comma-separated list of options for a declared
 particle,  and it may include the following items:
\begin{itemize}
\item the first element in this list must be the full name of the particle,
   (e.g. {\tt electron} and {\tt photon} in our example.) 
      Full name is a string constant, so it 
        should be quoted if it contains blanks, etc.
\item \underline{\tt mass \sl param=value} defines the mass of the
particle.
Here {\sl param} is an identifier of a new parameter, which is used to
denote the mass; {\sl  value} is its value,  
 it has the same syntax as in the {\tt parameter} statement, comment for
 this new parameter being generated automatically. If this option omitted,
 the mass is assumed to be zero.
\item \underline{\tt width \sl param=value} declares the width of the
 particle. It has the same syntax as for {\tt mass } option.
\item \underline{\tt texname \sl string} declares the name to use in LaTeX
output. 
\item \underline{\tt atexname \sl string} set LaTeX name for antiparticle, 
by default it is  $\backslash$bar\{texname\}. 
\item \underline{\sl itype} is a type of index of some symmetry; 
one can use default index types for color $SU(3)$ group (see QCD 
example in Section 2). It is possible to use user-defined index types
 (see Section \ref{newgrp}) and Lorentz group indices {\tt vector,
 spinor, cspinor}. 
\item \underline{\tt left} or \underline{\tt right} says that the
massless spinor particle is an eigenstate of $(1-\gamma_5)/2$ or
 $(1+\gamma_5)/2$ projectors, so this fermion is left-handed or
right-handed.
\item \underline{\tt gauge} declares the vector particle as a gauge boson.
 This  option generates the corresponding ghosts and goldstone bosons names
 for the named particle (see below).
\end{itemize}

When a particle name is used in any expression (in Lagrangian terms), one
should remember that the first index is either a vector or a spinor
(if this particle is not a Lorentz scalar). Then the
 indices follow in the
same order as index types in the {\sl options} list. So, in the case of 
quark declaration (see the QCD example) the first index is spinor, and the
second one is color triplet. 

There are several functions taking particle name as an argument which
can be used in algebraic expressions. These functions are replaced with 
auxiliary particle names, which are generated automatically.
\begin{itemize}
\item Ghost field names in gauge theories are generated by the functions
 {\tt ghost(\sl name\tt) $\rightarrow$ '\sl name\/\tt .c'} and
 {\tt ccghost(\sl name\tt) $\rightarrow$ '\sl name\/\tt .C'} (see for
 instance
 Table 1).
 Here and below
{\sl name} is the name of the corresponding gauge boson. 
\item Goldstone boson field name in the t'Hooft-Feynman gauge is generated
 by
the function  {\tt gsb(\sl name\tt) $\rightarrow$ '\sl name\/\tt .f'}.
\item The function {\tt anti(\sl name\/\tt)} generates antiparticle name
for the  particle {\sl name}. 
\item The name for a charge conjugate spinor particle 
$\psi^c=C\bar{\psi}{}^T$ is generated by the function 
 {\tt cc(\sl name\/\tt) $\rightarrow$ '\sl name\/\tt .c'}. Charge 
conjugate fermion has the same indices types and ordering, however index
of  {\tt spinor} type is replaced by the index type {\tt cspinor} and
 vice versa.
\item {\tt vev(\sl expr\tt\/)} is used in the Lagrangian for vacuum
 expectation values.
Function {\tt vev} ensures that {\tt deriv*vev(\sl expr\/\tt)} is zero.
In other words, {\tt vev} function forces LanHEP to treat {\sl expr} as
 a scalar particle
which will be replaced by {\sl expr} in Feynman rules.
\end{itemize}

\subsection{Specials}

Besides parameters and particles other indexed
such as $\gamma$-matrices, group structure constants, etc may appear in the
Lagrangian terms. 
We refer such objects as {\it specials}. 

Predefined specials of the Lorentz group are:
\begin{itemize}
\item \underline{\tt delta(vector)} can be used for $g_{\mu\nu}$
\item \underline{\tt gamma} stands for the $\gamma^\mu$-matrices. It has three
    indices of  {\tt spinor, cspinor} and {\tt vector} types.
\item \underline{\tt gamma5} denotes $\gamma_5$ matrix. It has two
    indices of  {\tt spinor} and {\tt cspinor}  types.
\item \underline{\tt moment} has one index of {\tt vector} type.
At the stage of Feynman rules generation this symbol is replaced by the
 particle moment.
\item \underline{\tt deriv}  is replaced 
by $-ip_\mu$, where $p_\mu$ is the particle moment. It has one {\tt vector}
 index. Note that {\tt deriv*A*B} means $(\partial A)B$, not $\partial (AB)$.
\item  For the derivative of a product, {\tt derivp(\it expr\tt)} can be
  used. {\tt derivp(A*B)} is equivalent to {\tt deriv*A*B+A*deriv*B}.
\item \underline{\tt epsv} is an antisymmetric tensor
$\epsilon^{\mu\nu\rho\sigma}$ with 4 tensor indices.
\end{itemize}

Specials of the color $SU(3)$ group are:
\begin{itemize}
\item \underline{\tt lambda} denotes Gell-Mann $\lambda$-matrices. It has
 three
indices: {\tt color c3, color c3b} and {\tt color c8}.
\item \underline{\tt f\_SU3} is the $SU(3)$ structure constant. It has
 three indices
of {\tt color c8} type.
\item \underline{\tt eps\_c3} and \underline{\tt eps\_c3b} are antisymmetric
  tensors. The first has 3 indices of the type {\tt color c3}, and the second
  3 {\tt color c3b} indices.
\end{itemize}

Note that for specials the order of indices types is fixed. 
  
Users can  declare new specials with the help of a facility defined in 
Section~\ref{newgrp} to introduce user-defined indices types. 

\subsection{Let-substitutions}

LanHEP allows the user to introduce new symbols and then substitute them
in Lagrangian terms by some
expressions. 
Substitution has the generic
form \begin{quote} {\tt let \sl name\/\tt =\sl expr. }\end{quote}
where {\sl name} is the identifier of newly defined object.
 The expression has the same structure as those in Lagrangian
terms, however here expression may have free (non-contracted) indices.

Typical example of using a
substitution rule is a definition of the QED  covariant derivative as
\begin{quote} {\tt let Deriv\^{}mu=deriv\^{}mu + i*ee*A\^{}mu. }\end{quote}
corresponding to $D_\mu=\partial_\mu + ig_eA_\mu$.

More complicated example is the declaration $\sigma^{\mu\nu}
\equiv i(\gamma^\mu\gamma^\nu-\gamma^\nu\gamma^\mu)/2$ matrices:
\begin{quote} 
{\tt let sigma\^{}a\^{}b\^{}mu\^{}nu = 
i*(gamma\^{}a\^{}c\^{}mu*gamma\^{}c\^{}b\^{}nu \\
\phantom{let sigma\^{}a\^{}b\^{}mu\^{}nu =} - gamma\^{}a\^{}c\^{}nu*gamma\^{}c\^{}b\^{}mu)/2.}
\end{quote}

Note that the order of indices types 
of new symbol is fixed by the declaration. So, first two indices of
{\tt sigma} after this declaration
are spinor and antispinor, third and fourth are vector indices.

\subsection{Arrays \label{taudef}}

LanHEP allows to define components of indexed objects. In this case,
contraction of indices will be performed as an explicit sum of products
of the corresponding components.

An object with explicit components  has to be written as
\begin{quote} {\tt \{\sl expr1, expr2 ..., exprN \tt \}\^{}\sl i
 }\end{quote}
where  expressions correspond to components. All indices of components
(if present) have to be written for each component, and the index 
numbering components  has to be written after closing curly bracket. 
Of course, all the components must have the same types of free
(non-contracted) indices.

Arrays are usually applied for the definition of multiplets and matrices
corresponding to broken symmetries. 

Typical example of arrays usage is a declaration of
electron-neutrino isospin doublet {\tt l1} (and antidoublet {\tt L1})
\begin{quote} {\tt let l1\^{}a\^{}I = \{ n1\^{}a, e1\^{}a\}\^{}I, 
L1\^{}a\^{}I = \{ N1\^{}a, E1\^{}a\}\^{}I. } \end{quote}
Here we suppose that {\tt n1} was declared as the spinor particle
 (neutrino), with the antiparticle name {\tt N1}.

Note that the functions {\tt anti} and {\tt cc} can be applied also to the
multiplets, so the construction
\begin{quote} {\tt let l1\^{}a\^{}I = \{ n1\^{}a, e1\^{}a\}\^{}I, 
L1\^{}a\^{}I = anti(l1)\^{}a\^{}I. } \end{quote}
is possible. 

Matrices can be represented as arrays which have other arrays as 
components.
However, it is more convenient to declare them with dummy indices, see
Section \ref{oimatr} (the same is correct for multiplets also).
 
It is possible also to use arrays directly in the Lagrangian, rather
 than  only in the declaration of let-substitution.

When LanHEP is launched, it has already declared some frequently used matrices.
They are:
\begin{itemize}
\item {\tt delta(\sl N\tt )} represents the unit matrix $\delta_{ab}$ of dimension $N$;
\item {\tt tau1, tau2, tau3} are $\tau$-matrices $\tau_1, \tau_2, \tau_3$:
\[
\tau^1 =  \left(\begin{array}{rr} 0 & 1 \\ 1 & 0 \end{array} \right), \;\;\;\;
\tau^2 =  \left(\begin{array}{rr} 0 & -i \\ i & 0 \end{array} \right), \;\;\;\;
\tau^3 =  \left(\begin{array}{rr} 1 & 0 \\ 0 & -1 \end{array} \right);
\]
\item {\tt tau} is a vector $\vec{\tau}=(\tau_1, \tau_2, \tau_3)$;
\item {\tt taup} and {\tt taum} are matrices $\tau^\pm=(\tau^1\pm i\tau^2)/
\sqrt{2}$;
\item {\tt taupm} is a vector $(\tau^+, \tau^3, \tau^-)$;
\item {\tt eps} is the antisymmetrical tensor
                 $\varepsilon$ ($\varepsilon^{123}=1$);
\item {\tt Tau1, Tau2, Tau3} are generators of $SU(2)$ group adjoint
representation (3-dimensional analog of $\tau$-matrices)
 $T^1, T^2, T^3$
 with commutative relations $[T_i,T_j]=-i\epsilon_{ijk}T_k$: 
\[
T^1= \frac{1}{\sqrt{2}} \left(\begin{array}{rrr} 0 & -1 & 0 \\
 -1 & 0 & 1 \\ 0 & 1 & 0 \end{array} \right), \;\;\;\;
T^2= \frac{1}{\sqrt{2}}  \left(\begin{array}{rrr} 0 & i & 0 \\
 -i & 0 & -i \\ 0 & i & 0  \end{array} \right), \;\;\;\;
T^3=\left(\begin{array}{rrr} 1 & 0 & 0 \\ 0 & 0 & 0 \\ 0 & 0 & -1
  \end{array} \right);
\]
\item {\tt Taup} and {\tt Taum} corresponds to $T^\pm=(T^1\pm iT^2)/
\sqrt{2}$;
\item {\tt Taupm} is a vector $\vec{T}=(T^+, T^3, T^-)$.
\end{itemize}

\subsection{Two-component notation for fermions}

It is possible to write the Lagrangian using two-component notation
for fermions. The connection between two-component and four-component 
 notations is summarized by the following relations:
$$
 \psi=\left(\begin{array}{c} \xi \\ \bar\eta \end{array} 
\right),\;\;\;\;\;\;
 \psi^c=\left(\begin{array}{c} \eta \\ \bar\xi \end{array}
\right),\;\;\;\;\;\;
 \bar\psi=\left(\begin{array}{c} \eta \\ \bar\xi \end{array}
\right)^T,\;\;\;\;\;\;
 \bar\psi^c=\left(\begin{array}{c} \xi \\ \bar\eta \end{array} \right)^T.
$$

If the user has declared a spinor particle {\tt p} (with antiparticle {\tt
P}), the LanHEP notation for its components is:

\begin{tabular}{rcl}
$\xi$ & $\rightarrow$ & {\tt up(p)}\\
$\bar\eta$ & $\rightarrow$ & {\tt down(p)}\\
$\eta$ & $\rightarrow$ & {\tt up(cc(p)) \it or \ \tt up(P)}\\ 
$\bar\xi$ & $\rightarrow$ & {\tt down(cc(p)) \it or \ \tt down(P)}
\end{tabular}

To use the four-vector $\bar\sigma^\mu$ one should use the statement
\begin{quote}
\hspace*{1cm}{\tt special sigma:(spinor2,spinor2,vector).}\end{quote}
Note that the {\tt sigma} object is not defined by default, thus the
above statement is required. It is 
possible also to use another name instead of {\tt sigma} (this object can be
recognized by LanHEP by the types of its indices).

LanHEP uses the following rules to convert the two-component fermions
to four-component ones (we use $P_{R,L}=(1\pm\gamma_5)/2$):\\[5mm]

\begin{tabular}{lrcl}
$\eta_1\xi_2=\bar\psi_1P_L\psi_2$ \phantom{xxxxx}& 
{\tt up(P1)*up(p2)} & $\rightarrow$ &  {\tt P1*(1-gamma5)/2*p2} \\
$\bar\eta_1\bar\xi_2=\bar\psi_1P_R\psi_2$ & 
{\tt down(P1)*down(p2)} & $\rightarrow$ & {\tt P1*(1+gamma5)/2*p2}\\
$\xi_1\xi_2=\bar\psi_1^cP_L\psi_2$ & 
{\tt up(p1)*up(p2)} & $\rightarrow$ &  {\tt cc(p1)*(1-gamma5)/2*p2} \\
$\bar\xi_1\bar\xi_2=\bar\psi_1P_R\psi_2^c$ & 
{\tt down(P1)*down(P2)} & $\rightarrow$ & {\tt P1*(1+gamma5)/2*cc(P2)}\\
$\xi_1\eta_2=\bar\psi_1^cP_L\psi_2^c$ & 
{\tt up(p1)*up(P2)} & $\rightarrow$ &  {\tt cc(p1)*(1-gamma5)/2*cc(P2)} \\
$\bar\xi_1\bar\eta_2=\bar\psi_1^cP_R\psi_2^c$ & 
{\tt down(p1)*down(P2)} & $\rightarrow$ & {\tt cc(p1)*(1+gamma5)/2*cc(P2)}\\
$\eta_1\eta_2=\bar\psi_1P_L\psi_2^c$ & 
{\tt up(P1)*up(P2)} & $\rightarrow$ &  {\tt P1*(1-gamma5)/2*cc(P2)} \\
$\bar\eta_1\bar\eta_2=\bar\psi_1^cP_R\psi_2$ & 
{\tt down(p1)*down(p2)} & $\rightarrow$ & {\tt
cc(p1)*(1+gamma5)/2*p2}\\[3mm]
$\bar\xi_1\sigma^\mu\xi_2=\bar\psi_1\gamma^\mu P_L\psi_2$ &
{\tt down(P1)*sigma*up(p2)} & $\rightarrow$ & {\tt P1*gamma*(1-gamma5)/2*p2}\\
$\bar\eta_1\sigma^\mu\eta_2=\bar\psi_1^c\gamma^\mu P_L\psi_2^c$ &
{\tt down(p1)*sigma*up(P2)} & $\rightarrow$ & 
{\tt cc(p1)*gamma*(1-gamma5)/2*cc(P2)}
\end{tabular}


\section{Lagrangian expressions}

When all parameters and particles necessary for the introduction of
a (physical) model are declared, one can enter Lagrangian terms with the help
of  the 
 {\tt lterm} statement:
\begin{quote} {\tt lterm \sl expr\tt . }\end{quote}

Elementary objects of expression are integers, identifiers of parameters,
particles, specials, let-substitutions, and arrays.

These elementary objects can be combined by the usual arithmetic operators
as 
\begin{itemize}
\item  {\sl expr1+expr2\/}  (addition),
\item  {\sl expr1--expr2\/}  (subtraction),
\item {\sl expr1*expr2\/}  (product),
\item {\sl expr1/expr2\/}  (fraction; here {\sl expr2} must be a product
 of integers and parameters),
\item {\sl expr1**N\/}  ({\sl N}th power of {\sl expr1}; {\sl N}
 must be an integer).
\end{itemize}
  One can use brackets '(' and ')' to force  the precedence
of operators. Note, that indices can follow only elementary objects
symbols, {\it i.e.} if {\tt A1} and {\tt A2} were declared as two vector
 particles then valid
expression for their sum is {\tt A1\^{}mu+A2\^{}mu}, rather than {\tt
(A1+A2)\^{}mu}.

\subsection{Where-substitutions}

More general form of expressions  involves {\it where-substitutions}:
\begin{quote} {\sl expr \tt where \sl subst}.\end{quote}
 
 In the simple form {\sl subst} is
 {\sl name\tt =\sl repl} or several constructions of this type separated by
comma ','. In the form of such kind each instance of identifier {\sl name}
 in {\sl expr} is
replaced by {\sl repl}. 

Note that in contrast to let-substitutions,
where-substitution does not create a new object. LanHEP simply replaces
 {\sl name} by 
{\sl repl}, and then processes the resulting expression. It means in
 particular
that  {\sl name} can not have indices, although it can denote an object
 with
indices:
\begin{quote} {\tt lterm F**2 
where F=deriv\^{}mu*A\^{}nu-deriv\^{}nu*A\^{}mu.}
\end{quote}
is equivalent to 
\begin{quote} {\tt lterm (deriv\^{}mu*A\^{}nu-deriv\^{}nu*A\^{}mu)**2.}
\end{quote} 

 The substitution rule introduced by 
the keyword {\tt where} is active only within the current {\tt lterm}
 statement.  

More general form of where-substitution allows to use several
{\sl name\/\tt =\sl repl} 
substitution rules separated by semicolon ';'. In this
case {\sl expr} will be replaced by the sum of expressions; each term
 in this sum
is produced by applying one of the substitution rules from
semicolon-separated list to the expression {\sl expr}.  This form is
 useful for writing
the Lagrangian where many particles have a similar interaction. 

For example, if {\tt u,d,s,c,b,t} are declared as quark names,
their interaction with the gluon may read as
\begin{quote} {\tt lterm gg*anti(psi)*gamma*lambda*G*psi where \\
\phantom{lterm gg*} psi=u; psi=d; psi=s; psi=c; psi=b; psi=t.} \end{quote}

The equivalent form is
\begin{quote} {\tt lterm  gg*U*gamma*lambda*G*u + gg*D*gamma*lambda*G*d +
 \\
\phantom{lterm} gg*C*gamma*lambda*G*c + gg*S*gamma*lambda*G*s + \\
\phantom{lterm} gg*T*gamma*lambda*G*t + gg*B*gamma*lambda*G*b. }
 \end{quote}

Where-substitution can also be used in {\tt let} statement. In this case
one should use brackets:
\begin{quote} {\tt let \sl lsub=(expr\/\ \tt where \sl wsub=expr1).}
\end{quote} 

\subsection{Adding Hermitian conjugate terms}

It is possible to generate hermitian conjugate terms automatically by
putting the symbol {\tt AddHermConj} to {\tt lterm} statement:
\begin{quote} {\tt
lterm \it expr\/ \tt + AddHermConj.}\end{quote}
Continuing the former example, one can write:
\begin{quote} {\tt
lterm a*H*H*h + b*H**3 + AddHermConj.}\end{quote}

Note, that the symbol {\tt AddHermConj} adds the hermitian conjugate 
expressions to all terms in the {\tt lterm} statement. It means in particular that
in the statement
\begin{quote} {\tt
lterm \it expr1\/ \tt + (\it expr2\/ \tt + AddHermConj).}\end{quote}
the conjugate terms are added to both {\it expr1\/} and {\it expr2\/}.
Thus, one should not place self-conjugate terms in the {\tt lterm}
statement where {\tt AddHermConj} present (or one should supply these
terms with $1/2$ factor).

\subsection{Using the superpotential formalism in the MSSM and its extensions}

In supersymmetric models (in particular, in the Minimal Supersymmetric 
Standard Model (MSSM) \cite{roziek}) one makes use of the superpotential ---
a polynomial $W$ depending on scalar fields $A_i$ (superpotential also can
be defined in terms of superfields, we do not consider this case).
Then, there is the contribution to the Lagrangian: Yukawa terms 
in the form 
$$ -\frac{1}{2}\left(\frac{\partial^2W}{\partial A_i \partial A_j}
\Psi_i \Psi_j + H.c.\right) $$  and  $F^*_iF_i$ terms, where 
$F_i= \partial W/ \partial A_i $ (for more details, see \cite{roziek}
and references therein). 

To use this formalism in LanHEP, one should define first the multiplets
of matter fields and then define the superpotential through the 
let-substitution statement.
The example of the MSSM with a single generation may read:
\begin{quote} {\tt keep\_lets W.\\
let W=eps*(mu*H1*H2+ml*H1*L*R+md*H1*Q*D+mu*H2*Q*U).
}\end{quote}
Here symbols {\tt H1, H2, L, R, Q, U, D} are defined somewhere else
as doublets and singlets in terms of scalar particles.

Note that before the definition of {\tt W} this symbol should
appear in the {\tt keep\_lets} statement. It is necessary to notify LanHEP
that let-substitutions (multiplets) at the definition of {\tt W} should not
be expanded. Without this statement, the representation used
by LanHEP for {\tt W} will not contain symbols
of multiplets but only the particles which were used at multiplets definition.

Since {\tt W} was declared in the {\tt keep\_lets} statement and contains the
symbols of multiplets, one can evaluate the variational derivative of {\tt W}
by one or two multiplets, e.g. {\tt df(W,H1)} or {\tt df(W,H1,L)}. Thus
the Yukawa terms may be written:
\begin{quote} {\tt lterm - df(W,H1,H2)*fH1*fH2 - ... + AddHermConj.}\end{quote}
Here {\tt fH1, fH2} are fermionic partners of corresponding multiplets.

To introduce the $F^*_iF_i$ terms one needs to declare the conjugate 
superpotential, e.g. {\tt Wc}, and write:
\begin{quote} {\tt lterm - df(W,H1)*df(Wc,H1c) - .... }\end{quote}
A better way is to use the function {\tt dfdfc(W,H1)} instead:
\begin{quote} {\tt lterm - dfdfc(W,H1) - .... }\end{quote}
The function evaluates the variational derivative, multiply it by the conjugate
expression and returns the result. Moreover, it can introduce auxiliary
fields to split vertices with 4 color particles (in the case of CompHEP output);
see Section \ref{imp} for more details.

\subsection{Generation of counterterms for calculation at 1 loop}
 
 Ultraviolet divergencies in renormalized quantum field theories
 are compensated by  renormalization of wave functions $\phi_i 
 \rightarrow \phi_i(1+\delta Z_i)$, masses $M_i \rightarrow M_i + 
 \delta M_i$, charges $e_i \rightarrow e_i+\delta e_i$, and may be 
 some other Lagrangian paramters. Such transformation of the 
 Lagrangian assumes changes in the Feynman rules.
 In particular vertices are changed due to the appearence of new terms --- 
 {\it counterterms}. It is possible to treat this transformation 
 at 1-loop level by means of LanHEP package. Currently this feature is not
 supported by CompHEP, but it can be used for 1-loop calculation 
 by FeynArts \cite{FeynArts} and FormCalc \cite{FormCalc} 
 (e.g. it is done in the SloopS \cite{sloops} package).
 
 In  1-loop approximation renormalization parameters appear in expressions
 only at first power, {\it i.e.} $\delta Z_i\delta Z_j=0$, where $\delta Z_i$ means 
all 
 renormalization parameters ($\delta e$, $\delta M_i$, $\delta Z_i$, ...).
  This should be done with the help
 of the following statement 
 \begin{quote} {\tt infinitesimal \it d\_Z1, d\_Z2, ..., d\_Zn.} \end{quote}
 It declares that a monomial should be omitted if it contains more than one 
 of parameters { \it d\_Z1, d\_Z2, ..., d\_Zn}.

  The statement
 \begin{quote} {\tt transform \it obj \tt -> \it expr\tt .} \end{quote}
 forces LanHEP to substitute symbol of a parameter or a particle, {\it obj}, 
 by the expression {\it expr} when this object appears in the Lagrangian
 expressions of the Lagrangian. 
 
 For example, if the electric charge $e$ in QED is renormalized than the 
 statement
 \begin{quote} {\tt transform ee -> ee+dZe. } \end{quote}
 makes LanHEP substitute each occurence of {\tt ee} symbol in further
 expressions by {\tt ee+dZe}. 
 
 The {\tt transform} statement is very similar to the {\tt let} 
  statement and 
 uses the 
 same syntax for expression {\it expr}. It has however two advantages. First,
 one does not need to introduce new name for transformed parameter or
 field. Second, if one has the LanHEP file for some physical model, it is
 enough to add counterterms statements into the input file just after  
 the statements declaring parameters and particles.

 An example for QED model with renormalization reads as:
 \begin{quote} {\tt
 parameter ee = 0.3133: 'Electric charge'.\\
 vector A/A:photon.\\
 spinor e1/E1:(electron, mass me=0.000511).\\
 infinitesimal dZee, dZme, dZe1l, dZe1r,  dZA.\\
 transform ee -> ee*(1+dZee), me -> me+dZme, \\
 \hspace*{1cm} A -> A*(1+dZA),\\
 \hspace*{1cm} e1 -> e1 + (dZel/2*(1-gamma5)/2+dZer/2*(1+gamma5)/2)*e1,\\
 \hspace*{1cm} E1 -> E1 + (dZel/2*(1+gamma5)/2+dZer/2*(1-gamma5)/2)*E1.\\
 lterm e*E1*(i*gamma*deriv+gamma*A+me)*e1. } \end{quote}

Sometimes it is necessary to block the shifts set by {\tt transform} 
statement in some parts of the Lagrangian (depending on the specific
renormalization scheme). This
can be done by {\tt BlockTransf} option:
\begin{quote} {\tt 
option BlockTransf=0.\\
\sl statements...\\
option BlockTransf=1.} \end{quote} 

Another option {\tt InfiOrder} can be used in the same way to set the maximal
power of infinitesimal parameters in the Lagrangian expressions. By default,
it is set to 1. It is also possible to define high order renormalization
parameters by applying {\tt **\sl N} in the declaration, like:
 \begin{quote} {\tt infinitesimal \it d\_Z12**2.} \end{quote}

If one use the two-component fermion notation, the shifts for the upper
and the lower components of a 4-component fermions must be defined.
This is done automatically if the shift for 4-component fermion is defined
as above for the electron, with left and right projectors. A special case
concerns Majorana fermions, e.g. neutralino in MSSM. The shift should be defined
here in a following way:
\begin{quote} {\tt
spinor n/n:(neutralino, mass Mn).\\
infinitesimal dZn/dZnc.\\
transform n -> n + (dZn/2*(1-gamma5)/2+dZnc/2*(1+gamma5)/2)*n.} \end{quote}
The second statement declares renormalization parameter {\tt dZn} and 
its complex conjugate {\tt dZnc}.

\subsection{Constructing the ghost Lagrangian}

The ghost Lagrangian can be constructed by means of the BRST formalism.
To use it, the user should first define the BRST
transformations for fields (see Section \ref{brst}).

 The gauge-fixing Lagrangian reads as:
$${\cal L}_{GF} = G^-G^+ + \frac{1}{2}|G^Z|^2 + \frac{1}{2}|G^\gamma|^2,$$
 where $G^i$ ($i=\pm,\gamma,Z$) are gauge-fixing functions.
 The ghost Lagrangian ensures the BRST invariance of the entire
 Lagrangian and can be written as (\cite{ghconstr})
 $${\cal L}_{Gh} = -\bar c^i \delta_{BRST} G^i + \delta_{BRST} \tilde{\cal L}_{Gh}.$$
 where $\delta_{BRST} \tilde{\cal{L}}_{Gh}$ is an overall function, which is 
 BRST-invariant (at 1 loop we can set this to zero).
 
So, for the photon and $G^\gamma = (\partial \cdot A)$, the LanHEP code
for gauge-fixing and ghost terms reads as:
\begin{quote}{\tt
let G\_A = deriv*A.\\
lterm  -G\_A**2/2.\\
lterm  -'A.C'*brst(G\_A).}\end{quote}
Here the {\tt brst} function is used to get BRST-transformation
of the specified expression. 

The inverse BRST transformation can also be used. One can declare
the transformations for the fields by means of {\tt brsti\_transform}.
 The function {\tt brsti(\it expr\tt)} can be used in {\tt lterm}
statements.


\section{Omitting indices}

Physicists usually do not write all possible indices in the Lagrangian
 terms.
LanHEP also allows a user
to omit indices. This feature can simplify the introduction of the
expressions and makes them more readable.
Compare two possible forms:
\begin{quote} {\tt lterm E1\^{}a*gamma\^{}a\^{}b\^{}mu*A\^{}mu*e1\^{}b. }
 \end{quote}
corresponding to $g_e \bar e_a(x) \gamma^\mu_{ab} e_b(x) A_\mu(x)$, and
\begin{quote} {\tt lterm E1*gamma\^{}mu*e1*A\^{}mu. }\end{quote}
corresponding to $g_e \bar e(x) \gamma^\mu e(x) A_\mu(x)$). 
Furthermore, while physicists usually write vector indices explicitly in 
the formulas, in LanHEP  vector indices also can be omitted:
\begin{quote} {\tt lterm -i*ee*E1*gamma*e1*A. }\end{quote}
Generally speaking, when the user omits indices in the expressions, LanHEP
 faces two  problems: which
indices were omitted  and
how to contract indices. 

\subsection{Restoring the omitted indices}

When the indexed object  is declared
 the corresponding set of indices is assumed.
 Thus, if the quark {\tt
q} is declared as
\begin{quote} {\tt spinor q:('some quark', color c3). }\end{quote}
its first index is spinor and the second one belongs to the {\tt color c3}
type.
If both indices are omitted in some expression, LanHEP generates
them in the correspondence to  order ({\tt spinor, color c3}).
However, if only one index is written, for example in
the form  {\tt q\^{}a}, LanHEP has to recognize
whether the index {\tt a} is of {\tt color c3} or of  {\tt spinor}
types.

To solve this problem LanHEP looks up the {\it list of indices
omitting order}.
By default this list is set to \begin{quote}
{\tt [spinor, color c3, color c8, vector]} \end{quote}

The algorithm to restore omitted indices is the following. First, LanHEP
assumes that the user has omitted
indices which belong to the first type (and corresponding antitype) from
 this
list. Continuing with our example with particle {\tt q}
 one can
see that since this particle is declared having one {\tt spinor} index (the
first type  in the list) LanHEP checks whether the number of
indices declared for this object without {\tt spinor} index equals the
number of indices written explicitly by user. In our example (when user
has written {\tt q\^{}a}) this is true. In the following LanHEP concludes 
that the user omitted the {\tt spinor} index and that  explicitly written
 index is
of {\tt color c3} type.

In other cases, when the supposition fails if the user has omitted indices
 of the 
first type in the {\it list of indices omitting order}, LanHEP goes to
the second step. It assumes that user
has omitted indices of first two
types from this list. If this assumption also fails, LanHEP 
assumes that the user has omitted indices of first three types in the list and
 so on. 
At each step LanHEP subtracts the number  of  indices of
these types assumed to be omitted from the full number of indices declared
 for the
object, and checks
whether this number of resting indices equals the number of 
explicitly written indices.  If LanHEP fails when the {\it list of indices
omitting order} is completed, error message is returned by the program.

Note that if the user would like to omit indices of some type, he must 
omit 
all indices of this type (and antitype) as well as  the
indices of all types which precede in the {\it list of indices omitting
order}.

For example, if object {\tt Y} is declared  with one {\tt spinor},
two {\tt vector} and three {\tt color c8} indices, than 
\begin{itemize}
\item the form {\tt Y\^{}a\^{}b\^{}c\^{}d\^{}e\^{}f} means that the user
 wrote 
all the indices explicitly;
\item the form {\tt Y\^{}a\^{}b\^{}c\^{}d\^{}e} means that the user
 omitted {\tt spinor} 
index and wrote {\tt vector} and {\tt color c8} ones;
\item the form {\tt Y\^{}a\^{}b} means that the user omitted {\tt spinor}
 and {\tt color c8}
indices and wrote only two {\tt vector} ones;
\item the form {\tt Y} means that the user omitted all indices;
\item all other forms, involving different number of written indices, are
 incorrect.
\end{itemize}


One could change the {\it list of indices omitting order
types} by the statement {\tt SetDefIndex}. For example, for default
setting it looks like
\begin{quote} {\tt SetDefIndex(spinor, color c3, color c8, vector).
 }\end{quote}
Each argument in the list is a type of index.

\subsection{Contraction of restored indices}

A dummy index can be contracted only with some another dummy index.
LanHEP expands the expression and restores indices in each monomial.
LanHEP reads objects in
the monomial  from the left to the right and checks whether the
restored indices are present. If such index appears LanHEP looks for the
 restored index of the appropriate type at the next objects. Note, that the
program does not check  whether the object with the first restored index
 has another
restored index of the appropriate type. Thus, if {\tt F} is declared 
as let-substitution for the
strength tensor of electromagnetic field (with two vector indices) then
expression {\tt F*F} (as well as {\tt F**2}) after processing dummy
 indices
turns to the implied form {\tt \tt F\^{}mu\^{}nu*\tt F\^{}mu\^{}nu} rather
than {\tt \tt F\^{}mu\^{}mu*\tt F\^{}nu\^{}nu}.

This algorithm makes the contractions to be sensitive on the order of
 objects in
the monomial. Let us look again at the QED example.
Expression {\tt E1*gamma*A*e1} (as well as {\tt A*e1*gamma*E1})
leads to correct result where the vector index of photon is contracted with the
same index of $\gamma$-matrix, spinor index of electron is contracted with
antispinor index of $\gamma$-matrix and antispinor index of
positron is contracted with spinor index of $\gamma$-matrix. However the
expression {\tt E1*e1*A*gamma} leads to the wrong form {\tt
E1\^{}a*e1\^{}a*A\^{}mu*gamma\^{}b\^{}c\^{}mu}, because the first
 antispinor 
index after the electron belongs to positron.
Spinor indices of {\tt gamma} stay free (non-contracted) since no more
 objects with
appropriate indices (so, LanHEP will report an error since
Lagrangian term is not a scalar).

Note, that in the vertex with two $\gamma$-matrices the situation is more 
ambiguous. Let's look at  the
term corresponding to the electron anomalous magnetic moment
$\bar e(x) (\gamma^\mu\gamma^\nu-\gamma^\nu\gamma^\mu)e(x)F_{\mu\nu}$.
The correct LanHEP expression is 
\begin{quote} {\tt e1*(gamma\^{}mu*gamma\^{}nu
- gamma\^{}nu*gamma\^{}mu)*E1*F\^{}mu\^{}nu} \end{quote}
Here vector indices can't
be omitted, since it lead to the contraction of vector indices of 
$\gamma$-matrices. One can see also that the form 
\begin{quote} {\tt e1*E1*(gamma\^{}mu*gamma\^{}nu
- gamma\^{}nu*gamma\^{}mu)*F\^{}mu\^{}nu} \end{quote} will correspond to 
the expression $\bar e(x) e(x) Tr(\gamma^\mu\gamma^\nu)F_{\mu\nu}$. 
Here LanHEP does see a scalar Lorentz-invariant expression in the Lagrangian
term, so it has no reason to report an error. 

These examples mean that the user
should clearly realize how the indices will be restored and
contracted, or (s)he has to write all indices explicitly.

\subsection{Let-substitutions}

Another problem arises when the dummy indices stay free, this is the case for
 the {\tt let} statement. 
LanHEP allows only two ways to avoid ambiguity in the order of indices
types:  either the user
specifies all the indices at the name of new symbol and free indices in
 the corresponding
expression, or he should omit all free indices.

In the latter case the order of indices types is defined following the order of
free dummy indices in the first monomial of the expression. For
example
if {\tt A1} and {\tt A2} are vectors and {\tt c1} and {\tt c2} are spinors,
the statement \begin{quote} {\tt let d=A1*c1+c2*A2.} \end{quote}
declares a new object {\tt d} with two indices, the first is a vector index 
and the second  is a spinor index according to their order in the monomial
{\tt A1*c1}. Of course, each monomial in the expression must have the same 
 type of free indices.

\subsection{Arrays \label{oimatr}}

The usage of arrays with dummy indices allows us to define matrices
conveniently.  For example, the declaration of
$\tau$-matrices (defined in Section \ref{taudef}) 
can be written as
\begin{quote} {\tt
let tau1\ = \{\{0,\hphantom{-}1\}, \{\hphantom{-}1,\hphantom{-}0\}\}. \\
let tau2 = \{\{0,\hphantom{-}i\}, \{-i,\hphantom{-}0\}\}. \\
let tau3 = \{\{1,\hphantom{-}0\}, \{\hphantom{-}0,-1\}\}. } \end{quote}
One can see that in such way of declaration a matrix is written
"column by column".

The declaration of objects with three `explicit' indices can
 be done using the
objects already defined. For example, when $\tau$-matrices are defined as
before, it is easy
to define the vector  $\vec{\tau}\equiv(\tau_1, \tau_2, \tau_3)$ as
\begin{quote} {\tt let tau = \{tau1, tau2, tau3\}. } \end{quote}
The object {\tt tau} has three indices, first pair selects the element of
the matrix, while the matrix itself is selected by the third index, {\it i.e.}
{\tt tau\^{}i\^{}j\^{}a} corresponds to $\tau^a_{ij}$.

On the other hand, the declaration of structure constants of a group 
is more complicated.
Declaring such an object one should bear in mind that omitting indices
implies that in a sequence of components the second index of an object 
is changed after the full cycle of the first index, the third index is
 changed
after the full cycle of the second one, etc.
For example, a declaration of the antisymmetrical tensor
 $\varepsilon^{abc}$
can read as
\begin{quote} {\tt
          let eps =   \{\{\{0,0,0\}, \{0,0,-1\}, \{0,1,0\}\}, \\
\hphantom{let eps = } \{\{0,0,1\}, \{0,0,0\}, \{-1,0,0\}\}, \\
\hphantom{let eps = } \{\{0,-1,0\}, \{1,0,0\}, \{0,0,0\}\}\}.
  } \end{quote}
One can easily see that the components  are listed here in the
 following order:
\begin{quote}
$\varepsilon^{111}, \varepsilon^{211}, \varepsilon^{311},
 \varepsilon^{121},
\varepsilon^{221}, \varepsilon^{321}, \varepsilon^{131},
\; ... \;\varepsilon^{233}, \varepsilon^{333}$.\end{quote}
The declaration of more complex objects such as $SU(3)$ structure constants
can be made in the same way.

\section{Checking the correctness of the model}

\subsection{Checking electric charge conservation}

LanHEP can check whether the introduced vertices satisfy electric charge
conservation law. This option is available, if the user declares some
 parameter
to denote elementary electric charge (say, {\tt ee} in QED example),
and then indicates, which particle is a photon  by the statement
\begin{quote} {\tt SetEM(\sl photon, param).}\end{quote}
So, in example of Section 1 this statement could be
\begin{quote} {\tt SetEM(A,ee).} \end{quote}

Electric charge of each particle is determined by analyzing its
 interaction 
with the photon. LanHEP
checks whether the sum of electric charges of particles in each vertex
equals zero. Command line option {\tt -v-charges} allows to display 
the electric charge detected.

\subsection{Testing Hermitian conjugate terms}

LanHEP is able to check the correctness
of the hermitian conjugate terms in the Lagrangian. To do this, user
should use the statement:
\begin{quote} {\tt CheckHerm. }\end{quote}

For example, let us consider the following (physically meaningless) input
file, where a charged scalar field is declared and cubic terms are introduced:
\begin{quote} {\tt
scalar h/H.\\
parameter a,b,c.\\
lterm a*(h*h*H+H*H*h)+b*h*h*H+c*H*H*h + h**3.\\
CheckHerm. 
}\end{quote}

The output of LanHEP is the following (here 2 is the symmetry factor):
\begin{verbatim}
CheckHerm: vertex (h, h, h): conjugate (H, H, H) not found.
CheckHerm: inconsistent conjugate vertices: 
    (H, h, h)             (H, H, h) 
      2*a          <->       2*a
      2*b          <->   (not found)
   (not found)     <->       2*c
\end{verbatim}

If the conjugate vertex is not found, a warning message is printed.
If both vertices are present but they are inconsistent, more detailed
output is provided. For each couple of the incorrect vertices LanHEP outputs 
3 kinds of monomials: 1) those which are found in both vertices (these 
monomials are correctly conjugated), 2) the monomials found only in first vertex,
 and 3) the monomials found only in the second vertex. 
 
\subsection{Probing the kinetic and mass terms of the Lagrangian}

 LanHEP allows user to examine whether the mass sector of
the Lagrangian is consistent\footnote{In CompHEP, the information
about particles propagators is taken from the particle declaration, where 
particle mass and width (for Breit-Wigner's propagators) are provided.
Thus, the user is not obliged to supply kinetic and mass terms in {\tt lterm}
statements. Even if these terms are written, they do not affect the CompHEP
output.}. To do this, use the statement 
\begin{quote}{\tt CheckMasses.}\end{quote}
This statement must be put after all {\tt lterm} statements of the input file.

When the {\tt CheckMasses} statement is used, LanHEP creates the file
named {\tt masses.chk} (in the directory where LanHEP was started).
This file contains warning messages
if some inconsistencies are found:
\begin{itemize}
\item missing or incorrect kinetic term;
\item mass terms with mass different from the value specified at the particle
declaration;
\item off-diagonal mass terms.
\end{itemize}

Note that in more involved models like the MSSM the masses could depend on 
other parameters and LanHEP is not able to prove that expressions in actual
mass matrix and in parameters declared to be the masses of particles are 
identical. Moreover, it is often impossible to express the masses as formulas
written in terms of independent parameters. For this reason LanHEP evaluates
the expressions appearing in the mass sector numerically (based on the parameter
values specified by the user).

It is typical for the MSSM that some fields are rotated by unitary matrices
to diagonalize mass terms. In some cases, values of mixing matrices elements
can not be expressed by formulas and need to be evaluated numerically.
LanHEP can recognize the mixing matrix if the elements of this matrix were
used in {\tt OrthMatrix} statement. LanHEP restores and prints the mass 
original matrix before fields rotation by mixing matrix. 

The numerical check of correctness of the mass matrix diagonalization can be
done if LanHEP has access to the routine which calculates the masses and 
the rotation matrix. To provide this access, one should should make the 
dynamic library, e.g.
\begin{quote}{\tt gcc -shared -o diag.so diag.c}\end{quote}
and then specify this library as the third argument of 
{\tt external\_func} statement:
\begin{quote} { \tt external\_func(diag,3,'./diag.so')}.\end{quote}
Note that './' is necessary unless you put your file into system
shared library path.

\subsection{Checking BRST invariance \label{brst}}

LanHEP can check the BRST invariance (see the reviews in \cite{brst1,brst2})
 of the Lagrangian. First, the user should declare the BRST
transformations for the fields in the model by means of {\tt brst\_transform}
statement:
\begin{quote}{\tt brst\_transform \it field \tt -> \it expression\tt .
}\end{quote}

For example, the BRST transformation for the photon in Standard Model
$\delta A_\mu = \partial_\mu c^A + ie(W^+_\mu c^- - W^-_\mu c^+)$
can be prescribed by the statement:
\begin{quote}{\tt
brst\_transform  A  ->  deriv*'A.c'+i*EE*('W+'*'W-.c'-'W-'*'W+.c').
}\end{quote}
The file 'sm\_brst.mdl' in LanHEP distribution contains the code
for gauge and Higgs fields transformation corresponding to the
CompHEP implementation of the Standard model.

Since the transformations are defined, the statement
\begin{quote}{\tt CheckBRST.}\end{quote}
enables the BRST transformation for Lagrangian terms (so it should
be placed before the first {\tt lterm} statement). The output
is CompHEP or LaTeX file. Certainly,
if the Lagrangian is correct, the output files are empty. However
some expressions identical to zero could be not simplified and remain
in the output.

\subsection{Extracting vertices}
 
      New statement allows to extract a class of vertices into separate
       file. This feature is useful in checking large models with thousands
     of vertices, such as the MSSM. 
 	The statement has the form
 
 \begin{quote} {\tt SelectVertices(\it file, vlist \tt). } \end{quote}
 
      Here {\it file} is a file name to output selected vertices, and 
    {\it vlist} determines the class of vertices. The pattern {\it vlist}
     may  have two possible formats.
 
      In the first format {\it vlist} is a list of particles: 
 \begin{quote} {\tt [\it P1, P2, P3, ... Pn \tt] } \end{quote}
      All vertices consisting only of the listed particles {\it P1, P2,...}  
      are extracted. For example, the pattern
 \begin{quote} {\tt [A, Z, 'W+', 'W-', e1, E1, n1, N1] } \end{quote}
      selects vertices of leptons of first generation interaction with gauge 
 	fields and the self-interaction of gauge fields (we used here
      particles notation of CompHEP).
      
      The second format reads
 \begin{quote} {\tt [\it P11/P12../P1n, P21/.../P2m, ... \tt ] } \end{quote}
      Here are several groups of particles, each group is joined by slash '/'
 	symbol.
      Selected vertices have the  number of legs equal to the number of 
      groups in the list, each leg in the vertex corresponds
      to one group. For example,
 \begin{quote} {\tt [e1/E1/n1/N1, e1/E1/n1/N1, A/Z/'W+'/'W-'] } \end{quote}
      selects vertices with two legs corresponding to leptons, third
       leg is gauge boson. Thus, this pattern extracts vertices of 
      leptons interactions with gauge bosons. Note that the 
      vertices of  gauge bosons interaction are not selected in
      this case. If a group consists of one particle, it must be typed 
 	 twice to enable second format (e.g. {\tt [P1/P1, P2/P2, P3/P3]}
 	 for the vertex consisting of particles {\tt P1, P2, P3}.
      
      It is possible to specify some options in {\tt SelectVertices} 
statement.
      The {\tt WithAnti} option adds antiparticle names for each particle 
      in the list (in the first format) or in the each group (in the second
      format. The {\tt Delete} option removes selected vertices from the 
      LanHEP's internal vertices list, so these vertices are not written 
      in output file {\tt lgrng\it N\tt .mdl (lgrng\it N\tt .tex)}. 
      This could be useful if {\tt SelectVertices}
      statements are supposed to distribute all vertices which should be in 
      the model, than some 'unexpected' vertices can be found in output 
      {\tt lgrng\it N\tt .mdl} file.
 
      An example of lepton-gauge vertices selection could read as
 \begin{quote} {\tt      
 	SelectVertices( 'lept-gauge.mdl', 
 			[e1/n1, e1/n1, A/Z/'W+'], WithAnti, Delete).}
 \end{quote}

\section{Simplifying the expression for vertices}

\subsection{Orthogonal matrices}

If some parameters appear to be the elements of the orthogonal matrix such
as quark mixing Cabbibo-Kobayashi-Maskava matrix, one should declare them
  by the statement
\begin{quote} {\tt OrthMatrix( \{\{$a_{11}$, $a_{12}$, $a_{13}$\}, 
\{$a_{21}$, $a_{22}$, $a_{23}$\}, \{$a_{31}$, $a_{32}$, $a_{33}$\}\} ).} 
\end{quote}
where $a_{ij}$ denote the parameters. Such a declaration permits LanHEP
to reduce expressions which contain these parameters by taking into
account the properties of the orthogonal matrices. If the parameters 
are named as above, {\tt a11, a12, ... a33} one can use the short form:
\begin{quote} {\tt OrthMatrix(a\_\_ , 3).}  \end{quote} Here two underline '\_' 
characters stands for 2 digits, and the numer 3 sets the dimension of matrix.

Note that this statement has no relation to the arrays;
it just declares that these parameters $a_{ij}$ satisfy the correspondent
relations.
If necessary, one can declare further a matrix with these parameters
as components by means of the {\tt let} statement.

\subsection{Reducing trigonometric expressions}
 
 Some physical models, such as the MSSM 
 and general Two Higgs Doublet Model, involve 
 large expressions built up of trigonometric functions. In particlular, after
 the diagonalization of the Lagrangian in the Higgs sector, in the
 models mentioned above two angles, $\alpha$ and $\beta$, are introduced, thus 
 the Lagrangian may be written in LanHEP notation using the following
 definitions:
 \begin{quote} {\tt 
 parameter sa=0.5:'sinus alpha', \\
 \hspace*{2cm} ca=Sqrt(1-sa**2):'cosine alpha'.\\
 parameter sb=0.9:'sinus beta', \\
 \hspace*{2cm} cb=Sqrt(1-sb**2):'cosine beta'. }\end{quote}
 One can find in the output expressions like {\tt sa*cb+ca*sb} 
 $=\sin(\alpha+\beta)$ and much more complicated ones. To simplify the 
 output and 
 make it more readable the user can define new parameters, for example:
 \begin{quote} {\tt 
 parameter sapb=sa*cb+ca*sb:'sin(a+b)'.}\end{quote}
 In order to force LanHEP to substitute the expression {\tt sa*cb+ca*sb} by the 
 parameter {\tt sapb},
 one should use the statement
 \begin{quote} {\tt 
 SetAngle(sa*cb+ca*sb=sapb).} \end{quote}
 It is possible also to substitute an expressions by any polynomial
 involving paramters, for example,
 \begin{quote} {\tt 
 SetAngle((sa*cb+ca*sb)**2+(sa*cb-ca*sb)**2=sapb**2+samb**2).} \end{quote}
 where {\tt samb} should be previously declared as a parameter.
 Note, that LanHEP expands both expressions before setting the substitution
 rule. The left-hand side expression can also contain symbols defined by {\tt let}
 statement. In the Lagrangian expressions LanHEP expresses powers of 
 cosines, {\tt ca**\it N\/} and {\tt cb**\it N\/} through sines using
 recursively the formula 
 $\cos^N\alpha= \cos^{N-2}(1-\sin^2\alpha)$ to combine all similar terms.
  The same procedure is performed
 for the left-hand side expression in {\tt SetAngle} statement.
 
 If the substitution for one angle combination is defined, LanHEP 
 offers to define the other ones for expressions consisting of
 parameters used in previous {\tt SetAngle} statements. In our example,
 for all expressions consisting of parameters {\tt ca, cb, sa, sb},
 the following message is printed (e.g. for $\cos(\alpha+\beta)$):
 \begin{quote} {\tt
 Warning: undefined angle combination; use:\\
 \hspace*{2cm}	SetAngle(ca*cb-sa*sb=aa000). } \end{quote}
 Here {\tt aa000} is an automatically generated parameter.
 The message is printed only once for each expression. 
 This feature is disabled by default; to enable producing these messages,
 one should issue the statement
 \begin{quote} {\tt option UndefAngleComb=1. }\end{quote}
 
\subsection{Heuristics for trigonometric expressions}

LanHEP can apply several heuristic algorithms to simplify
these expressions. For each angle $\alpha$, the user should declare parameters
for  $\sin\alpha$, $\cos\alpha$, $\sin 2\alpha$, $\cos 2\alpha$, $\tan\alpha$.
Than one should use the {\tt angle} statement:
\begin{quote}{\tt
angle sin=\sl p1\tt, cos=\sl p2\tt, sin2=\sl p3\tt, cos2=\sl p4\tt,
tan=\sl p5\tt, texname=\sl name\tt .}\end{quote}
Here {\sl pN} --- parameter identifiers, {\sl name} --- LaTeX name for
angle, it is used to generate automatically LaTeX names for trigonometric
functions of this angle if these names are not set explicitly by 
{\tt SetTexName} statement. This statement should immediately follow 
the declaration of the parameters for $\sin\alpha$ and $\cos\alpha$.
Only the {\tt sin} and {\tt cos} options are mandatory. Other parameters
({\it i.e.} $\sin 2\alpha$, $\cos 2\alpha$, $\tan\alpha$) should be declared 
if these parameters are defined before $\sin\alpha$ and $\cos\alpha$.
If the former parameters will be declared in terms of latter ones,
they are recognized automatically and they need not appear in {\tt angle}
statements.

For example, the declaration for trigonometric functions of $\beta$ angle
in MSSM may read:
\begin{quote} {\tt
parameter tb=2.52:'Tangent beta'.\\
parameter sb=tb/Sqrt(1+tb**2):'Sinus beta'.\\
parameter cb=Sqrt(1-sb**2):'Cosine beta'.\\[2mm]
angle sin=sb, cos=cb, tan=tb, texname='$\backslash\backslash$beta'.\\[2mm]
parameter s2b=2*sb*cb:'Sinus 2 beta'.\\
parameter c2b=cb**2-sb**2:'Cosine 2 beta'. } \end{quote}
Here the parameters {\tt s2b} and {\tt c2b} are recognized automatically 
by LanHEP as $\sin 2\alpha$ and $\cos 2\alpha$ since they are declared 
in terms of {\tt sa, ca} parameters.

For a couple of angles ( $\alpha$, $\beta$ in this example) the user should
declare the parameters for $\sin(\alpha+\beta)$, $\cos(\alpha+\beta)$,
$\sin(\alpha-\beta)$, and $\cos(\alpha-\beta)$ to allow using all implemented
heuristics:
\begin{quote} {\tt
parameter sapb=sa*cb+ca*sb : 'sin(a+b)'.\\ 
parameter samb=sa*cb-ca*sb : 'sin(a-b)'.\\ 
parameter capb=ca*cb-sa*sb : 'cos(a+b)'.\\ 
parameter camb=ca*cb+sa*sb : 'cos(a-b)'. } \end{quote}
These parameters are recognized as trigonometric functions automatically,
by analysis of the right-hand side expressions.

It is possible to control the usage of heuristics by the statement:
\begin{quote} {\tt option SmartAngleComb=\sl N\tt.}\end{quote}
where {\sl N} is a number:
\begin{enumerate}
\item[0] heuristics are switched off;
\item[1] heuristics are switched on (by default);
\item[2] same as 1, prints the generated substitution rule if the simplified
expressions consists of more than 3 monomials;
\item[3] same as 1, prints all generated substitution rule.
\item[4] same as 1, prints all generated substitution rule and some 
      intermediate expressions (debug mode).
\end{enumerate}

The substitution rules are printed as {\tt SetAngle} statement and can
be used for manual improvement of expressions.

\subsection{Abbreviations}

The long expressions can be removed from the vertices and placed into
the parameters table. This trick allows to optimize significantly
the further work of matrix element computation both in memory consumption
and computation speed. To generate these new auxiliary parameters
(abbreviations) one should run LanHEP with the command line option 
{\tt -evl \sl num\/}. For the lengthly expression in a vertex
(more then {\sl num\/} monomials) an auxiliary variable is generated,
and the expression is stored in the parameters table.
 {\sl num=2\/} is recommended.

In case of FeynArts output, one should use {\tt -abbr} command
line option which produce optimal code for this program.

\section{Output}

Currently, LanHEP can generate Feynman rules in one of three formats:
CompHEP, FeynArts, LaTeX.

\subsection{CompHEP}
By default, LanHEP produces 4 files with tables for independent parameters
{\tt vars\sl N\tt .mdl}, derived ones {\tt func\sl N\tt .mdl},
particles {\tt prtcls\sl N\tt .mdl}, and vertices {\tt lgrng\sl N\tt .mdl},
where {\sl N} is the number specified in the {\tt model} statement.
To start working with CompHEP in the new model, one should copy these files into
the {\tt models/} subdirectory of the CompHEP working directory. The 
number {\sl N} should be next to the number of models already existing 
in this subdirectory.

\subsubsection{Modifying the particle table format}

New option allows to modify the format of the output particle 
table and to add new proprties (new columns in the table).
The new format is set by means of {\tt prtcformat} statement, for example,
the default CompHEP table format can be set by:
\begin{verbatim}
prtcformat fullname:' Full   Name ',
           name:' p ', 
           aname:' ap',
           spin2,color,mass,width, aux,
           texname:'  latex P name ',
           atexname:' latex aP name ' .
\end{verbatim}

Each column of the table is described by the entry {\sl prop:title=value},
where {\sl prop} is the name of particle property, {\sl title} is shown
in the title line, and {\sl value} is default value used if the property
of specific particle is not set. {\bf Important:} the width of {\sl title}
fixes the width of the table column, so it should be wide enough to contain
records for this property for any particle. If not specified, property name
is used as title, and blank space for default value.

There is a set of predefined properties (table columns) which exist in 
CompHEP by now plus the  electric charge of a particle:

\begin{itemize}
\item {\tt fullname} 'Long' particle name.
\item {\tt name} Particle name used in vertices table.
\item {\tt aname} Antiparticle.
\item {\tt spin2} Twice spin, integer 0---4.
\item {\tt mass}  Particle mass.
\item {\tt width} Particle width.
\item {\tt echarge} Particle electric charge, integer or ratio $N/3$.
                    The value is generated automatically if the
					{\tt CheckEM(\it photon, coupling\tt)} statement
					is used in the model file.
\item {\tt echarge3} Three times electric charge, integer.
\item {\tt color} Dimension of the color $SU(3)$ group representation (one of 1,3,8).
\item {\tt aux} Specify the particle as left or right fermion, boson in 
Feynman gauge (see CompHEP manual).
\item {\tt texname} LaTeX notation to be used for this particle. 
\item {\tt atexname} Same for antiparticle. If not set, {\tt  bar\{texname\} }
is used.
\end{itemize}

Besides these predefined properties, the user can introduce new ones.
One can add, say, PDG particle number to the table:
\begin{verbatim}
prtcformat fullname:' Full   Name ',
           name:' p ', 
           aname:' ap',
           spin2,color,mass,width, aux,
           pdg:'PDG ID',
           texname:'  latex P name ',
           atexname:' latex aP name ' .
\end{verbatim}
Then the pair {\sl prop value} can be written in the particle declaration
statement:
\begin{quote}
 {\tt scalar h:(higgs, mass Mh, pdg 123, width wh). } \end{quote}

Another statement {\tt prtcprop} can solve two problems. First, it can 
declare a property (or a few of them) which are not included in the
particle table, but they still can be used in particle declarations
for further use in the future. For example, one can declare {\tt pdg}
property:
\begin{quote}
 {\tt prtcprop pdg. } \end{quote}
and use it in the particle declarations even though CompHEP does not
support this property now for future extensions. In this case,
particle table format is not changed. Several properties can be listed,
separated by comma.

Another problem is to set a new property for all particles in the model.
There is no need to modify all particles declarations in the model, one
can use statement like this:
\begin{quote}
 {\tt prtcprop pdg:(h=123, g=124, f=125). } \end{quote}
In this example h,g,f are some particles and the numbers are values
of a property. Among the predefined properties, only the mass and width can be
assigned (or changed)  in this
way.

If some property is set to a number, a ratio, some particle or parameter 
name, then it can be used in the Lagrangian expressions as {\sl (prop particle)},
e.g. electric charge can be used to describe quark-photon interaction:
\begin{quote}
 {\tt lterm (echarge q)*EE*Q*gamma*q*A. } \end{quote}
Of course one have to choose beween using the declared charge in expressions 
and automatic detection via {\tt SetEM} statement.

\subsubsection{Lagrangian table setup}

The format and content of the Lagrangian (vertices) table can be tuned
by several options set by {\tt option} statement:
\begin{quote}
 {\tt option \sl opname\tt =\sl value\tt . } \end{quote}
and by some command line options.

\begin{itemize}
\item {\tt chepCFWidth} Width of Common Factor field, default is 50.
\item {\tt chepLPWidth} Width of Lorentz Part field, default is 600.
\item {\tt RemDotWithFerm} options (default value $1$) tells LanHEP
to replace $g_{\mu\nu}$ by $(\gamma_\mu\gamma_\nu+\gamma_\nu\gamma_\mu)/2$ 
in the vertices with fermions (CompHEP requirement). If this replacement is
not neccessary put this option to $0$.
\item {\tt ReduceGamma5} (default value $1$) removes $(1\pm \gamma_5)/2$
operators in the vertices with lefthand or righthand fermions (e.g. neutrinos)
replacing them by 0 or 1. Put this option to 0 to keep projector operators
in such vertices.
\item {\tt MultByI} (default value $0$ for CompHEP output, $1$ for LaTeX)
In CompHEP tables, common $i$ in the vertices is not shown, so imaginary
unit appears only in the vertices with pseudoscalars. Setting this option to 
$1$ allows to restore it, making ComHEP vertices the same as in textbooks.
\item {\tt WriteAll} (default 0) Setting this option to 1 makes LanHEP
write into CompHEP table all vertices: 2-legs vertices,
      1-legs ones (they can appear from incorrectly written Higgs
      potential) and also the vertices with more than 4 legs. The command
	  line option {\tt -allvrt} has the same effect.  
\item {\tt MaxiLegs} allows to limit the number of legs in the vertices
       produced when {\tt WriteAll} is active.
\item {\tt WriteColors} (default 0) Set this option to 1 to write color structure
         of vertices explicitely. Also the command line options {\tt -colors} can
		 be used.
 \end{itemize}

When the {\tt WriteColors} option is set to 1, color matrices and dot products
are written in the Lorentz Part, e.g. QCD plus quark-photon interactions
produces the following vertices file:
\begin{verbatim}
QCD
 Lagrangian 
P1   |P2   |P3   |P4   |>   Factor   <|> dLagrangian/ dA(p1) dA(p2) dA(p3)
G    |G    |G    |     |gg            |m2.p3*m1.m3*F(c1,c2,c3)
     |     |     |     |               -m1.p3*m2.m3*F(c1,c2,c3)
     |     |     |     |               +m3.p1*m1.m2*F(c1,c2,c3)
     |     |     |     |               -m2.p1*m1.m3*F(c1,c2,c3)
     |     |     |     |               -m3.p2*m1.m2*F(c1,c2,c3)
     |     |     |     |               +m1.p2*m2.m3*F(c1,c2,c3)
G.C  |G.c  |G    |     |-gg           |m3.p2*F(c1,c2,c3)
Q    |q    |G    |     |gg            |L(c1,c2,c3)*G(m3)
Q    |q    |A    |     |ee/3          |c1.c2*G(m3)
G    |G    |G    |G    |gg^2          |m1.m3*m2.m4*F(c1,c2,c0)*F(c3,c4,c0)
     |     |     |     |               -m1.m4*m2.m3*F(c1,c2,c0)*F(c3,c4,c0)
     |     |     |     |               +m1.m2*m3.m4*F(c1,c3,c0)*F(c2,c4,c0)
     |     |     |     |               -m1.m4*m2.m3*F(c1,c3,c0)*F(c2,c4,c0)
     |     |     |     |               +m1.m2*m3.m4*F(c1,c4,c0)*F(c2,c3,c0)
     |     |     |     |               -m1.m3*m2.m4*F(c1,c4,c0)*F(c2,c3,c0)
\end{verbatim}

Here q/Q is a quark, G -- gluon , G.c -- gloun ghost field, and A -- photon.
F stand for antisymmetric structure constants (D for symmetric ones), L --
Gell-Mann matrices. The Lorentz part is shown one monomial per line just to make
the expression fit into the page, usually there is one line per vertex (option 
{\tt chepBreakLines} set to 1 to made this effect).

\subsection{FeynArts format}

Feynman rules in FeynArts format are generated if the command line option
{\tt -feynarts} is used at LanHEP start. In this case 4 files are generated.
First, {\tt model\sl N\tt .gen} contains 'generic model', in particular 
the Lorentz structures which appears in the model. {\sl N} is the number
defined in the {\tt model} statement in the LanHEP input file. 
Note that default
generic model {\tt FeynArts/Models/Lorentz.gen} can not be used with 
LanHEP output. Second file, {\tt model\sl N\tt .mod} 'classes model' defines interaction
vertices. Two other files are used to compile FORTRAN program from FormCalc 
output. {\tt model\sl N\tt .h} contains declarations of parameters and 
common blocks,  
{\tt mdl\_ini\sl N\tt .F} serve to initialize parameters.

To use the model generated by LanHEP one should have in Mathematica
input file the statement like:
\begin{quote} {\tt
SetOptions[InsertFields,Model->model\sl N\tt , GenericModel->model\sl N\tt ]} \end{quote}
The process can be set, for example for electron--positron annihilation into
two photons :
\begin{quote} {\tt
process = \{prt["e1"],prt["E1"]\} -> \{prt["A"],prt["A"]\}} \end{quote}
Here particle names defined in the LanHEP model can be used.

When FORTRAN code is generated by FormCalc, in the output replace the file
{\tt model.h} by {\tt model\sl N\tt .h}, and in the file {\tt run.F}
find {\tt \#include "sm\_ini.F"} and replace {\tt sm\_ini.F} by 
{\tt mdl\_ini\sl N\tt .F} 

Note that FormCalc has some model specific definitions, which can be seen
in the end of {\tt FormCalc.m} file. They can change some parameters
of the model generated by LanHEP. The safest way is to create model satisfying
these definitions. In particular, for each particle mass (say {\tt MH})
one should declare (in the LanHEP input file) its square like
\begin{quote} {\tt
parameter MH2=MH**2. }\end{quote}

The command line option {\tt -abbr} allows to create abbreviations
for the interaction vertices, placing the expressions from vertices into
parameters definition and making the model more compact and
increasing the speed of calculation of the matrix element.

If one has in mind 1-loop computations, renormalization parameters should be
defined. One can put the expression to be used by FormCalc in the 
{\tt infinitesimal} statement, e.g. for electric charge, weak angle and
tadpole counterterm one can write:
\begin{quote} {\tt infinitesimal   dEE= -(dZAA - SW/CW*dZZA)/2,\\
\phantom{xxxxxxxx}	dSW= CW\^{}2/SW/2*(dMZsq/MZ\^{}2 - dMWsq/MW\^{}2), \\
\phantom{xxxxxxxx}	dTad='-ReTilde[SelfEnergy[prt["H"] -> \{\}, -1]]'.}\end{quote}
The expression should be in quotes '' if it does not satisfy LanHEP syntax
('some[thing]' and '\{\}' do not). If the expression is not provided for a parameter, some default value is
used as in the standard FeynArts {\tt SM.mod} file, thus for the Standard Model
only these 3 parameters have to be set explicitely. 
Note that the old sign definition (as in FormCalc 4.1) is used for vector 
self-energies in the default scheme.


This output mode was tested with FeynArts 3.2 and FormCalc 4.1.

\subsection{LaTeX output}

LanHEP generates LaTeX output instead of CompHEP model files if the user 
set {\tt -tex} in the command line to start LanHEP.
Three files are produced: {\tt 'vars\sl N\/\tt .tex'}, 
{\tt 'prtcls\sl N\/\tt .tex'} and {\tt 'lgrng\sl N\/\tt .tex'}.
The first file contains names of parameter used in physical model
and their values. The second file describes the particles,
together with propagators derived from the vertices.

The last file lists introduced vertices. LanHEP uses Greek letters
$\mu, \nu, \rho$... for vector indices, letters $a,b,c$... for 
spinor ones and $p,q,r$... for color indices (and for indices of other
groups, if they were defined).

It is possible to inscribe names for particles and parameters
to use them in LaTeX output. It can be done by the statement
\begin{quote} {\tt SetTexName([ \sl ident=texname, ... \tt]).}\end{quote}
Here {\sl ident\/} is an identifier of particle or parameter, and 
{\sl texname\/} is string constant containing LaTeX command.
Note, that for introducing backslash '$\backslash$' in quoted string
 constant
one should type it twice: '$\backslash\backslash$'.

For example, if one has declared neutrino with name  {\tt n1} (and
name for antineutrino {\tt N1}) than the statement
\begin{quote} {\tt SetTexName([n1='$\backslash\backslash$nu\^{}e',
N1='$\backslash\backslash$bar\{$\backslash\backslash$nu\}\^{}e']).
}\end{quote}
makes LanHEP to use symbols $\nu^e$ and $\bar{\nu}^e$ for neutrino 
and antineutrino in LaTeX tables.

The vertices table can be tuned by the following command line options:
\begin{itemize}
\item[ ] \underline{\tt -texLines \sl num\/} Set number of lines in LaTeX
tables to {\sl num}. After the specified number of lines, LanHEP continues
writing current table on the next page of LaTeX the output. Default
 value is 40.
\item[ ] \underline{\tt -texLineLength \sl num\/} Controls  width of the
 Lagrangian
table.  Default value is 35, user can vary table width by changing this
parameter.
\item[ ] \underline{\tt -nocdot} removes $\backslash$cdot commands
between the parameters symbols in the LaTeX output. This option is
useful when all parameters have prescribed LaTeX names.
\item[ ] \underline{\tt -frc} If {\tt -tex\/} option is set, forces LanHEP
 to split
4-fermion and 4-color vertices just as it is made for CompHEP files.
\end{itemize}

\section{Declaration of new index types and indexed objects \label{newgrp}}

\subsection{Declaring new groups}

Index type is defined by two  
keywords\footnote{The exception is Lorentz group, corresponding indices
types are defined by a single keyword.}:   {\it group  name}
and   {\it representation name}. Thus, color triplet 
index type {\tt color c3} has group name {\tt color} and representation
name {\tt c3}.
 
LanHEP allows user to introduce new group names
by the  {\tt group} statement:
\begin{quote} {\tt group  \sl gname.}
\end{quote} Here {\sl gname } is a string constant,
 which becomes the name of newly declared group.

Representation names for each group name must be declared by the statement
\begin{quote} {\tt repres \sl gname\tt :(\sl rlist) }
\end{quote}
where  {\sl rlist } is a comma-separated list of representation names
declaration for the already declared group name {\sl gname}.
Each such declaration has the form either {\sl rname} or 
{\sl rname\tt /\sl crname}.  In the first case  index
which belongs to the {\sl gname rname} type can be contracted with
another index of the same type; in the second case index of 
{\sl gname rname} type can be contracted only with an index of 
{\sl gname crname} type.

For example, definition for color $SU(3)$ group with fundamental,
conjugate fundamental and adjoint representations looks like:
\begin{quote} { \tt group color:SU(3). \\ repres color:(c3/c3b,c8). }
\end{quote}
So, three index types can be used: {\tt color c3,
color c3b, color c8}. The contraction of these indices is allowed by
pairs ({\tt color c3, color c3b}) and ({\tt color c8, color c8}) indices.

\subsection{Declaring new specials}

Specials with indices of user-defined types can be declared by
means of
{\tt special} statement: \begin{quote} { \tt special \sl name\/\tt :(\sl
ilist\tt). } \end{quote}
Here {\sl name} is the name of new symbol, and {\sl ilist} is a
comma-separated list of indices types. For example,
Gell-Mann matrices can be defined as (although color
group and its indices types are already defined):
\begin{quote} {\tt special lambda:(color c3, color c3b, color c8).
 }\end{quote}

To define Dirac's $\gamma$-matrices one can use the command
\begin{quote} {\tt special gamma:(spinor, cspinor, vector). }\end{quote}

\subsection{Arrays}

Array, {\it i.e.} the object with explicit components, can also have the
user-defined type of index. In this case generic form of such object
 is
\begin{quote} {\tt \{ \sl expr1, expr2 ... ,exprN ; itype \} }\end{quote}
where $N$ expressions {\sl expr1 ... exprN} of $N$ components are separated
 by comma,
and {\sl itype} is an optional index type.
 If {\sl itype} is
omitted LanHEP uses default group name {\tt wild} and index type
{\tt wild \sl N}, where {\sl N} is a number of components in the array.

\section{Splitting the vertices with 4 colored particles \label{imp}}

The CompHEP Lagrangian tables do not describe  explicitly the color
structure of a vertex. If color particles are present in the vertex,
the following implicit contractions are assumed (supposing $p,q,r$
are color indices of particles in the vertex):
\begin{itemize}
\item $\delta_{pq}$ for two color particles $A^1_p$, $A^2_q$;
\item $\lambda_{pq}^r$ for three particles, which are color triplet,
antitriplet and octet;
\item $f^{pqr}$ for three color octets.
\end{itemize}
Other color structures are forbidden in CompHEP.

So, to introduce the 4-gluon vertex 
$f^{pqr}G_\mu^qG_\nu^rf^{pst}G_\mu^sG_\nu^t$ one should
split this 4-legs vertex into 3-legs vertices 
$f^{pqr}G_\mu^qG_\nu^rX_{\mu\nu}^p$:

{\scriptsize
\linethickness{0.5pt}
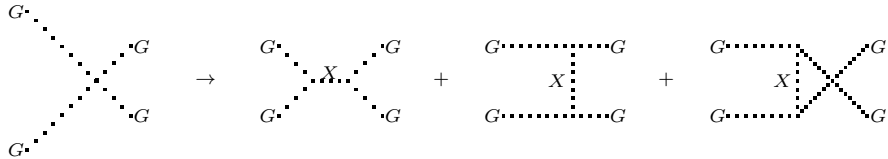
\begin{center}
\begin{picture}(62,77)(0,0)
\put(9.9,70.2){\makebox(0,0)[r]{$G$}}
\multiput(10.4,70.2)(3.3,-3.3){9}{\rule[-0.5pt]{1.0pt}{1.0pt}}
\put(9.9,17.8){\makebox(0,0)[r]{$G$}}
\multiput(10.4,17.8)(3.3,3.3){9}{\rule[-0.5pt]{1.0pt}{1.0pt}}
\put(51.1,57.1){\makebox(0,0)[l]{$G$}}
\multiput(36.5,44.0)(3.3,3.3){5}{\rule[-0.5pt]{1.0pt}{1.0pt}}
\put(51.1,30.9){\makebox(0,0)[l]{$G$}}
\multiput(36.5,44.0)(3.3,-3.3){5}{\rule[-0.5pt]{1.0pt}{1.0pt}}
\end{picture} \ 
\begin{picture}(25,77)(0,0)
\put(7,43){$\rightarrow$}
\end{picture}
\begin{picture}(62,77)(0,0)
\put(9.9,57.1){\makebox(0,0)[r]{$G$}}
\multiput(10.4,57.1)(3.3,-3.3){5}{\rule[-0.5pt]{1.0pt}{1.0pt}}
\put(9.9,30.9){\makebox(0,0)[r]{$G$}}
\multiput(10.4,30.9)(3.3,3.3){5}{\rule[-0.5pt]{1.0pt}{1.0pt}}
\put(30.2,47.1){\makebox(0,0){$X$}}
\multiput(23.5,44.0)(3.3,0.0){5}{\rule[-0.5pt]{1.0pt}{1.0pt}}
\put(51.1,57.1){\makebox(0,0)[l]{$G$}}
\multiput(36.5,44.0)(3.3,3.3){5}{\rule[-0.5pt]{1.0pt}{1.0pt}}
\put(51.1,30.9){\makebox(0,0)[l]{$G$}}
\multiput(36.5,44.0)(3.3,-3.3){5}{\rule[-0.5pt]{1.0pt}{1.0pt}}
\end{picture} \
\begin{picture}(15,77)(0,0)
\put(2,43){$+$}
\end{picture} 
\begin{picture}(62,77)(0,0)
\put(9.9,57.1){\makebox(0,0)[r]{$G$}}
\multiput(10.4,57.1)(3.3,0.0){9}{\rule[-0.5pt]{1.0pt}{1.0pt}}
\put(51.1,57.1){\makebox(0,0)[l]{$G$}}
\multiput(36.5,57.1)(3.3,0.0){5}{\rule[-0.5pt]{1.0pt}{1.0pt}}
\put(34.9,44.0){\makebox(0,0)[r]{$X$}}
\multiput(36.5,57.1)(0.0,-3.3){9}{\rule[-0.5pt]{1.0pt}{1.0pt}}
\put(9.9,30.9){\makebox(0,0)[r]{$G$}}
\multiput(10.4,30.9)(3.3,0.0){9}{\rule[-0.5pt]{1.0pt}{1.0pt}}
\put(51.1,30.9){\makebox(0,0)[l]{$G$}}
\multiput(36.5,30.9)(3.3,0.0){5}{\rule[-0.5pt]{1.0pt}{1.0pt}}
\end{picture} \ 
\begin{picture}(15,77)(0,0)
\put(2,43){$+$}
\end{picture} 
\begin{picture}(62,77)(0,0)
\put(9.9,57.1){\makebox(0,0)[r]{$G$}}
\multiput(10.4,57.1)(3.3,0.0){9}{\rule[-0.5pt]{1.0pt}{1.0pt}}
\put(64.5,57.1){\makebox(0,0)[l]{$G$}}
\multiput(36.5,57.1)(1.65,-1.65){17}{\rule[-0.5pt]{1.0pt}{1.0pt}}
\put(34.9,44.0){\makebox(0,0)[r]{$X$}}
\multiput(36.5,57.1)(0.0,-3.3){9}{\rule[-0.5pt]{1.0pt}{1.0pt}}
\put(9.9,30.9){\makebox(0,0)[r]{$G$}}
\multiput(10.4,30.9)(3.3,0.0){9}{\rule[-0.5pt]{1.0pt}{1.0pt}}
\put(64.5,30.9){\makebox(0,0)[l]{$G$}}
\multiput(36.5,30.9)(1.65,1.65){17}{\rule[-0.5pt]{1.0pt}{1.0pt}}
\end{picture} \ 
\end{center}
}

Here the field $X_{\mu\nu}^p$ is a Lorenz tensor and color octet,
and this field has constant propagator. If gluon name in CompHEP is 
{\tt 'G'}, the name {\tt 'G.t'} is used for this tensor particle;
its indices denoted as {\tt 'm\_'} and {\tt 'M\_'} ({\tt '\_'} is
the number of the particle in table item).

The described transformation is performed by LanHEP automatically and 
transparently for the user. Each vertex containing 4 color particles
is split to 2 vertices which are joined by an automatically generated
auxiliary field.

\subsection{Optimization in case of supersymmetry}

The same technique is applied in the MSSM where more vertices with 
4 color particles appear: vertices with 2 gluons and 2 squarks and vertices
with 4 squarks. However, the large amount of vertices with 4 squarks
requires many auxiliary fields, which can easily break CompHEP limitations
on the particles number. It is possible however to reduce significantly
the number of vertices and auxiliary fields if one introduce auxiliary
fields at the level of multiplets.

The vertices with 4 squarks come from $DD$ and $F^*F$ terms.
For example, there is  the term $\frac{1}{2}D^a_G D^a_G$ in the Lagrangian,
$$ D^a_G = g_s(Q_i^*\lambda^a Q_i + D^*\lambda^aD + D^*\lambda^aD), $$
where $Q,D,U$ are squarks multiplets, and $\lambda$ is Gell-Mann matrix.
Instead of evaluating this expression and that splitting all vertices
independently, one can introduce one color octet auxiliary field $\xi^a$
and write this Lagrangian term as $D^a_G \xi^a$.

Other $DD$ terms contain both color and colorless particles.
Thus, the term $D_A^iD_A^i$ with
$$ D_A^i= g_1(Q^* T^i Q + L^* T^i L + H_1^* T^i H_1 + H_2^* T^i H_2),$$
can be represented as 
$$ g_1(Q^* T^i Q)\xi^i + 
g_1^2(Q^* T^i Q)(L^* T^i L + H_1^* T^i H_1 + H_2^* T^i H_2) +
g_1^2(L^* T^i L + H_1^* T^i H_1 + H_2^* T^i H_2)^2$$
where $\xi^i$ is the triplet of auxiliary fields. This terms can also 
be written in the another form:
$$g_1(Q^* T^i Q + L^* T^i L)\xi^i +
g_1^2(Q^* T^i Q + L^* T^i L) (H_1^* T^i H_1 + H_2^* T^i H_2) +
g_1^2(H_1^* T^i H_1 + H_2^* T^i H_2)^2,$$
where all vertices with 4 scalars (except vertices with Higgs particles)
are splitted. Although the latter splitting is not mandatory, it can 
reduce  significantly the amount of vertices.

The similar technique is applicable to the $F^*F$ terms, with 
the transformation $F_i^*F_i\rightarrow F_i^*\xi_i^* + F_i\xi_i$.

Thus, we distinguish two types of vertices splitting: splitting at
multiplet level and splitting at vertices level. Note that splitting 
the vertices with two gluons and two squarks must be done at vertices
level after combining the similar terms, otherwise they would contain
the elements of squark mixing matrices.

The vertices splitting at multiplet level is implemented in LanHEP
mainly for MSSM needs. The first case refers to $DD$ terms. The user should
declare several let-substitutions and then put in {\tt lterm} statement
the squared sum:
\begin{quote} {\tt
let a1=g*Q*tau*q/2,\\
\phantom{let} a2=g*L*tau*l/2,\\
\phantom{let} a3=g*H1*tau*h1/2,\\
\phantom{let} a4=g*H2*tau*h2/2.\\
lterm   - ( a1 + a2 + a3 + a4 ) ** 2 / 2.}\end{quote}

In this case LanHEP looks for the square of the sum of several 
let-substitution symbols, each containing two color or merely scalar 
particles. If such an expression is found, it is replaced as in the previous
formulas.

The vertices splitting in $F^*F$ terms is performed by {\tt dfdfc} function
(see previous section). After taking the variational derivative the monomials
with two color or scalar particles (except Higgs ones) are 
multiplied by auxiliary fields, thus mediating the vertices with 4 color
(scalar) particles.

The multiplet level vertices splitting is controlled by the statement
\begin{quote}{\tt option SplitCol1=\sl N.}\end{quote}
where {\sl N} is a number:
\begin{enumerate}
\item[-1] remove all vertices with 4 color particles from Lagrangian;
\item[0] turn off multiplet level vertices splitting;
\item[1] allows vertices splitting with 4 color multiplets;
\item[2] allows vertices splitting with any 4 scalar multiplets
 except Higgs ones (more generally, any multiplets containing vev's).
\end{enumerate}
The value of this option can be set to different values before
executing different {\tt lterm} statements.

The vertices level splitting is performed after combining similar
terms of the Lagrangian. This splitting can be controlled by the statement
\begin{quote}{\tt option SplitCol2=\sl N.}\end{quote}
where {\sl N} is a number:
\begin{enumerate}
\item[0] disable vertex level splitting;
\item[1] enable vertex level splitting (only for vertices with 4 color
particles).
\end{enumerate}

For CompHEP output, the default value is 2 for {\tt SplitCol1} and
1 for {\tt SplitCol2}. For FeynArts and LaTeX output, default value is 0 for both 
options.

\subsection{Splitting vertices manually}

Sometimes it is necessary to split the vertex by hands. Usually 
this happens if the automatically generated vertices appears to be 
non-hermitian. Then one should write the Lagrangian using this 
auxiliary fields explicitely. An auxiliary particle can be generated
by means of {\tt AuxPrt(\sl spin, color, charged\tt )} function. Here {\sl spin}
is one of 0, 1, or 2, {\sl color} is one of 1, 3, 8, and {\sl charged}
is 0 for truly neutral particle, and 1 otherwise. For charged particle 
one should write like
\begin{quote}{\tt 
let a=AuxPrt(0,1,1), A=anti(a). }\end{quote}
Note that {\tt anti(AuxPrt(...))} will generate another auxiliary particle
and thus is wrong.

\section{Installation and running LanHEP}

To install LanHEP on your computer, you should get the archive file 
from the WWW-page { \tt http://theory.sinp.msu.ru/\~{}semenov/lanhep.html}.
Unpack the archive:
\begin{quote} {\tt gunzip lhep300.tar.gz \\ tar xf lhep300.tar }\end{quote}
The archive contains the directory {\tt lanhep} with C source files. 
If you do not have gcc compiler, you should tune the makefile,
change 'gcc' to the compiler which you have.
To create the executable file (called {\tt lhep}), go to {\tt lanhep} 
directory and type 
\begin{quote} {\tt make }\end{quote}
When LanHEP is complied, remove the source files by typing 
\begin{quote} {\tt make clean}\end{quote}
The archive also contains the subdirectory {\tt mdl} with startup file and
examples for several physical models. Add the directory containing LanHEP
to your {\sl PATH} environment variable. Then LanHEP can be started from
any other directory, it can find automatically files in the {\tt mdl} directory.

As mentioned above, LanHEP can read the model description 
from the input file prepared by the user.
To start LanHEP write the command
\begin{quote} {\tt lhep \sl filename options} \end{quote}
where the possible {\sl options} are described in the next section.
If the {\sl filename} is omitted, LanHEP prints
it's prompt and waits for the keyboard input. In the last case,
user's input is copied into the file {\tt lhep.log } and can be inspected
in the following. To finish the work with LanHEP,
type {\tt 'quit.'} or simply press {\tt \^{}D }
(or {\tt \^{}Z} at MS DOS computers).

Possible options, which can be used in the command line to start LanHEP
are:
\begin{itemize}
\item[ ] \underline{\tt -v} Verbose output: to provide information 
about processing the model. {\tt -vv} and {\tt -vvv} provide more information.
\item[ ] \underline{\tt -OutDir \sl directory\/} Set the directory where
 output files
will be placed.
\item[ ] \underline{\tt -InDir \sl directory\/} Set the default directory
 where to
search files, which included by {\tt read} and {\tt use} statements.
\item[ ] \underline{\tt -c3} produces the output files for CompHEP
version 3x.xx, which has slightly different format than 4x.xx.
\end{itemize}

\section*{Acknowlegements}

This work was supported in part by the French ANR project ToolsDmColl Blan07-2-194882.
I would like to thank N.\ Baro, F.\ Boudjema and D.\ Temes for collaboration
on the LanHEP interface to FeynArts through the SloopS project.

\eject

\tableofcontents

\end{document}